\documentclass[submission,copyright,creativecommons]{eptcs}
\providecommand{\event}{ACCAT 2012} 
\usepackage[normalem]{ulem}
\title{Category Theory and Model-Driven Engineering: %
From Formal Semantics to Design Patterns and Beyond}%
\author{Zinovy Diskin$^{1,2}$ \qquad  \qquad
            Tom Maibaum$^{1}$
 \institute{%
$^{1}$Network for Engineering of Complex Software-Intensive Systems
 for Automotive Systems (NECSIS),\\
McMaster University, Canada\\[5pt]
$^{2}$Generative Software Development Lab, \\
University of Waterloo, Canada \\ 
}
\email{diskinz@mcmaster.ca \qquad\qquad tom@maibaum.org}
} 
\usepackage
{graphicx}\graphicspath{{Figures/}}
\usepackage{rotating}
\usepackage{lscape}
\usepackage{epsfig}
\usepackage{color}
\usepackage{float}
\usepackage{amsfonts}
\usepackage{amsmath}
\usepackage{amstext}
\usepackage{amssymb}
\usepackage{mathrsfs}
\usepackage{array}
\usepackage{latexsym}
\usepackage{xspace}
\usepackage{wrapfig}
\usepackage{subfig}
\usepackage{multirow}

\newlength{\goupA}
\newlength{\goupB}
\newlength{\goupC}
\newcounter{Razdel}[subsection]
\newcommand\beginRazd[1]%
{

\par
\noindent%
\refstepcounter{Razdel}%
{\bfseries%
\small \thesubsection.\arabic{Razdel} #1} 
}

\newcommand\mysubsubsection[1]{\par\bigskip
\beginRazd{\scshape{#1}}\par\medskip\par\noindent
}

\newcounter{RazDel}[section]
\newcommand\beginRazD[1]
{

\par
\noindent\refstepcounter{RazDel}%
{\bf \thesection.\arabic{RazDel} #1} 
}

\newcommand\EndRazD{\par\smallskip}
\newcommand\EndRazd{\par\smallskip}

\newenvironment{razD}[2]{\beginRazD{{
                                      \bfseries #1}}}{\EndRazD}

\usepackage{diagrams}
\newarrowhead{bt}\blacktriangleright\blacktriangleright\blacktriangleright\blacktriangleright
\newarrowtail{bt}\blacktriangleleft\blacktriangleleft\blacktriangleleft\blacktriangleleft
\newarrowmiddle{bar}\rtbar\ltbar\dtbar\utbar
\newarrowmiddle{|}\vert\vert--
\newarrowmiddle{||}\Vert\Vert==
\newarrowmiddle{=}==\Vert\Vert
\newarrowtail{=}==\Vert\Vert
\newarrowhead{l}\langle\langle\langle\langle 
\newarrowhead{r}\rangle\rangle\rangle\rangle 
\newarrowfiller{=}==\Vert\Vert
\newarrowfiller{d}\cdot\cdot\cdot\cdot
\newarrowmiddle{x}****
\newarrowmiddle{b}\bullet\bullet\bullet\bullet
\newarrowtail{b}\bullet\bullet\bullet\bullet
\newarrowmiddle{3}\equiv\equiv\vfthree\vfthree
\newarrowfiller{3}\equiv\equiv\vfthree\vfthree
\newarrowtail{3}\equiv\equiv\vfthree\vfthree
\newarrowtail{<=}\Leftarrow\Rightarrow{\@cmex7E}{\@cmex7F}
\newarrowfiller{bold}{\bf-}{\bf-}{\bf|}{\bf|}  
\newarrowfiller{o}{\hho}{\circ}{\circ}{\circ}  
\newarrowmiddle{>}\rtla\ltla\dtla\utla
\newarrowmiddle{d}\cdot\cdot\cdot\cdot

\newarrow{To}----{>}
\newarrow{Mapsto}b---{>}
\newarrow{Maps}b----
\newarrow{ID}33333

\newarrow{EQ}=====
\newarrow{NRelto}--+-{->}
\newarrow{Relto}--b-{->}
\newarrow{BSpanto}--b-{->} 
\newarrow{Embed}>---{->}
\newarrow{Dashto}{}{dash}{}{dash}{>} 
\newarrow{RDiagto}3333{r}
\newarrow{RDiagderto}{}{3}{}{3}{r}
\newarrow{LDiagto}3333{l}
\newarrow{LDiagderto}{}{3}{}{3}{l}
\newarrow{Mapsderto}{b}{dash}{}{dash}{>}
\newarrow{Into}C---{>}
\newarrow{Indashto}C{dash}{}{dash}{>}
\newarrow{Derinto}C{dash}{}{dash}{>}
\newarrow{Congruent}33333
\newarrow{Cover}----{blacktriangle}
\newarrow{Monic}{>}--->
\newarrow{Isoto}{>}---{triangle}
\newarrow{ISA}===={=>}
\newarrow{Isato}===={=>}
\newarrow{Doubleto}===={=>}
\newarrow{ClassicMapsto}{|}--->
\newarrow{Entail}{|}----
\newarrow{PArrow}{o}--->
\newarrow{MArrow}----{>>}
\newarrow{MPArrow}{o}---{>>}
\newarrow{Ogogoto}oooo{->}
\newarrow{Metato}--3->
\newarrow{MetaMapto}{|}-3->
\newarrow{OneToMany}+---{o}
\newarrow{HalfDashTo}{}{dash}{}-{>>}
\newarrow{DLine}=====
\newarrow{Line}-----
\newarrow{Tline}33333
\newarrow{Dashline}{}{dash}{}{dash}{}
\newarrow{Dotline}{}{o}{}{o}{}
\newarrow{Curlyto}{curlyvee}--->           
\newarrow{Bito}<--->
\newarrow{Bito}{<}---{>}
\newarrow{Bidito}{<}==={>}
\newarrow{Bitrito}{bt}333{bt}
\newarrow{Corrto}<--->
\newarrow{Dercorrto}{<}{dash}{}{dash}{>}

\newarrow{Instofto}{curlyvee}...{>}
\newarrow{Instofderto}{curlyvee}{dash}{}{dash}{->}
\newarrow{Updto}----{>}
\newarrow{Derupdto}{}{dash}{}{dash}{>}
\newarrow{Mchto}----{>}
\newarrow{Dermchto}{}{dash}{}{dash}{>}
\newarrow{Viewto}----{>->}
\newarrow{Derviewto}{}{dash}{}{dash}{>>}
\newarrow{Viewto}===={=>}
\newarrow{Hetmchto}===={=>}
\newarrow{Derviewto}{=}{}{=}{}{=>}
\newarrow{Idleto}===={=>}
\newarrow{Deridleto}{=}{}{=}{}{=>}

\newlength{\StatArrBody}
\newlength{\NodeFrameThickness}
\newlength{\cellH}
\newlength{\cellW}

\newcommand\xx{\!\!}
\newcommand\xxx{\!\!\!}
\newcommand\xxxx{\xxx\xx\,}

\newcommand\seTilearrow[1]{
 \lefteqn{~~~{:}#1}{\searrow\xxx\xx\searrow}
 }%
\newcommand\swTilearrow[1]{%
 \lefteqn{~~~{:}#1}{\swarrow\xxx\xx\swarrow}
}%
\renewcommand\seTilearrow[1]{
 \lefteqn{~~~~{:}#1}{\searrow\xxxx\xxx\searrow}
 }%
\renewcommand\swTilearrow[1]{%
 \lefteqn{~~~~{:}#1}{\swarrow\xxxx\xxx\swarrow}
}%

\makeatletter
  \def\fps@figure{tp}
\makeatother

\newcounter{defCounter}
\newenvironment{defin}[1][]{
\refstepcounter{defCounter}
\begin{trivlist}
\item[\hskip \labelsep {\bfseries Definition \arabic{defCounter} (#1) }]}%
{\end{trivlist}}

\newcommand\syncon{synchronization}

\newcommand\figref[1]{Fig.~\ref{#1}}

\newcommand\sectref[1]{Sect.~\ref{#1}}

{\end{list}}

{\begin{equation}}{\end{equation}}

\newcommand\syncn{synchronization}

\newcommand{\hgs}{heterogeneous}

\newcommand\eg{e.g.}
\newcommand\ie{i.e.}

\newcommand\etc{{\em etc}}
\newcommand\etal{{\em et al}}

\newcommand\dolan
{\mbox{$\Delta $}}

\newcommand\mathsymbol[1]{\mbox{$#1$}}
\newcommand\world
{\mathsymbol{\cal W}}
\newcommand\cmter
{\mathsymbol{\cal C}}

\newcommand\formalObj[1]{\mbox{\sffamily #1}}

\newcommand\informalObj[1]{\mbox{\textit #1}}

\newcommand\FormalInterpret[1]{\mbox{$\formalObj{i}_f$}}
\newcommand\SubstantInterpret[1]{\mbox{$\informalObj{i}_s$}}

\renewcommand\eg{e.g.}

\newcommand\setcat{\mbox{$\setcatmath$}}
\renewcommand\setcat{\mbox{$\categoryname{Sets}$}}

\newcommand\tymap[1]{\ensuremath{t_{#1}}}
\newcommand\cartliarr[2]{\mbox{$#1{\star}#2$}}

\newcommand\cartliob[2]{\mbox{$#1{\bullet}#2$}}

\renewcommand\cartliob[2]{\mbox{$#1{\upharpoonright}_{#2}$}}

\renewcommand\cartliarr[2]{\mbox{$\overline{{#1}_{#2}}$}}

\newcommand\bilar[3]{\mbox{$#1\!:#2\leftrightarrow #3$}}%
\newcommand\flar[3]{\mbox{$#1\!:#2\rightarrow #3$}} 
\newcommand\proflar[3]{\mbox{$#1\!:#2\nrightarrow #3$}} 

\newcommand\bflar[3]{\mbox{$#1\!:#3\leftarrow #2$}} 
\newcommand\bdlar[3]{\mbox{$#1\!:#3\Leftarrow #2$}} 

\newcommand\dlar[3]{\mbox{$#1\!:#2\Rightarrow #3$}} 

\newcommand\spacenam[1]{\ensuremath{\mathbf{#1}}}

\newcommand\spaA{\spacenam{A}}
\newcommand\spaspaA{\ensuremath{\spacenam{A}^\bullet}}
\newcommand\spaspaB{\ensuremath{\spacenam{B}^\bullet}}

\newcommand\spaB{\spacenam{B}}

\newcommand\lanename[1]{\textbf{#1}}

\newcommand\rlane{\lanename{r}}

\newcommand\modcat{\catname{Mod}}

\newcommand\mmodcat{\catname{MMod}}

\newcommand\catname[1]{\mbox{\bfseries\itshape #1}}

\renewcommand\setcat{\catname{Set}}

\newcommand\ikeymap[1]{\mbox{$\mathbf{i}_{\mathrm{key}}$}}

\newcommand\diagopername[1]{\textsf{#1}}

\newcommand\ide[2]{\ensuremath{\mathsf{id}_{{}_{#1}}^{#2}}}
\renewcommand\ide[2]{\ensuremath{\mathsf{id}_{{}_{#1}}{\!#2}}}

\newcommand\falt{\ensuremath{\diagopername{fAln}}}
\newcommand\balt{\ensuremath{\diagopername{bAln}}}

\newcommand\fppg{\mbox{\diagopername{fPpg}}}
\newcommand\bppg{\mbox{\diagopername{bPpg}}}

\newcommand\get[1]{\mbox{\diagopername{Get}$^{#1}$}}
\renewcommand\put[1]{\mbox{\diagopername{Put}$^{#1}$}}

\newcommand\idppglaw{\lawname{IdPpg}}

\newcommand\fbfppglaw{\lawname{fbfPpg}}
\newcommand\bfbppglaw{\lawname{bfbPpg}}
\newcommand\ppgppglaw{\lawname{PpgPpg}}

\newcommand\altaltlaw{\lawname{AlnAln}}
\newcommand\idaltlaw{\lawname{IdAln}}

\newcommand\lawname[1]{\textsf{{#1}}}
\newcommand\putputlaw{\lawname{PutPut}}

\newcommand\idputlaw{\lawname{IdPut}}
\newcommand\putgetlaw{\lawname{PutGet}}
\newcommand\getputlaw{\lawname{GetPut}}
\newcommand\getputgetlaw{\lawname{GetPutGet}}

\newcommand\alens{l} 

\newcommand\spanl[1]{\mbox{$\overleftrightarrow{\alens}$}}

\newcommand\spanrla[1]{\mbox{$\overleftrightarrow{\rlane}$}}

\newcommand\mymarginnew[1]{ 
           \marginpar{{\flushleft\textit{\footnotesize #1}}}
           }
\newcommand\zdmargin[1]{\mymargin{ZDmarginNote:\;#1}}

\newcommand\yxmargin[1]{\mymargin{YX:\;#1}}

\def\modify#1#2#3#4#5{%
{\small\underline{\sf{#1}}}
{{#2}}
{\mymarginnew{#3}}
{#4}
{{#5}}
}%

\newcommand\new[2]{\modify{}{}{\sf{ZD #2}}{}{{\color{blue}#1}}}
\newcommand\newok[2]{#1}
\newcommand\newno[2]{\modify{}{}{}{{}}}

\newcommand\del[2]{\sout{{\color{red}#1}}\mymarginnew{\color{red}#2}}
\newcommand\delok[2]{}{} 

\newcommand{\zdmodiffy}[3]{\modify{}{\sout{\color{red}#1}}{ZD: \color{blue}#3}{\interArrow}{\color{blue}#2}}
\newcommand{\zdmodifyok}[2]{{#2}}

\renewcommand\modify[5]{{#5}}
\renewcommand\new[2]{{#1}}{}
\renewcommand\del[2]{}
\renewcommand{\zdmodiffy}[3]{{#2}}

\renewcommand\zdmargin[1]{}
\renewcommand\yxmargin[1]{}

\newarrow{Updto}----{>}
\newarrow{Derupdto}{}{dash}{}{dash}{>}
\newarrow{Mchto}----{>}
\newarrow{Dermchto}{}{dash}{}{dash}{>}
\newarrow{Uto}----{>}
\newarrow{Uderto}{}{dash}{}{dash}{>}
\newarrow{Mto}----{>}
\newarrow{Mderto}{}{dash}{}{dash}{>}

\newcommand\fwk{framework}
\newcommand\dbox[1]{\fbox{$#1$}}
\newcommand\aln{alignment}

\newcommand\GG{\ensuremath{\categoryname{Gra}}}

\newcommand\projfun{\ensuremath{\pmb p}}
\newcommand\pfi[1]{\ensuremath{\pmb p_{#1}}}
\newcommand\pfiq{\pfi{\qmon}}

\newcommand\qmon{\ensuremath{\mathsf{Q}}}
\newcommand\qmond{\ensuremath{\mathsf{Q}_{\mathrm{def}}}}

\def\titlerunning{Category Theory and Model Driven Engineering}
\def\authorrunning{Z. Diskin \& T. Maibaum}

\begin{document}
\maketitle
\begin{abstract}
   There is a hidden intrigue in the title. CT is one of the most
abstract mathematical disciplines, sometimes nicknamed "abstract
nonsense". MDE is a recent trend in software development, industrially
supported by standards, tools, and the status of a new \del{candidate for
the}{just to make it shorter} "silver bullet". Surprisingly, categorical
patterns turn out to be directly applicable to mathematical modeling of
structures appearing in everyday MDE practice. Model merging,
transformation, synchronization, and other important model management
scenarios can be seen as executions of categorical specifications.

Moreover, the paper aims to elucidate a claim that relationships between
CT and MDE are more complex and richer than is normally assumed for
"applied mathematics". CT provides a toolbox of design patterns and
structural principles of real practical value for MDE. We will present
examples of how an elementary categorical arrangement of a model
management scenario \zdmodiffy{makes explicit}{reveals}{?}
deficiencies in the architecture of modern tools automating the scenario.
\end{abstract}
%

\renewcommand\mysubsubsection[1]{\par\medskip\noindent\textbf{#1}}

\section{Introduction}

There are several well established applications of category theory (CT) in
theoretical computer science; typical examples are programming language
semantics and concurrency. Modern software engineering (SE) seems to
be an essentially different domain, not obviously suitable for theoretical
foundations based on abstract algebra. Too much in this domain
\zdmodifyok{seems}{appears to be} ad hoc and empirical, and the rapid
progress of open source and collaborative software development,
service-oriented programming, and cloud computing far outpaces  their
theoretical support. Model driven (software) engineering (MDE) conforms
to this description as well: {the diversity of modeling languages and
techniques successfully resists all attempts to classify them in a precise
mathematical way}, and model transformations and operations
--- MDE's heart and soul --- are  {an area of a diverse
experimental activity} based on surprisingly weak (if any) semantic
foundations.

In this paper we claim that theoretical underpinning of modern SE could
(and actually quite naturally) be based on CT. The chasm between SE and
CT can be bridged, and MDE appears as a \new{``golden cut''}{is it good?
the standard phrase is golden ratio}, in which an abstract view of SE
realities and concrete interpretations of categorical abstractions merge
together: SE $\rightarrow$ MDE $\leftarrow$ CT.  The left leg of the
cospan is extensively discussed in the MDE literature (see \cite{bran08}
and references therein); prerequisites and challenges for building the right
leg are discussed in the present paper. Moreover, we aim to elucidate a
claim that relationships between CT and MDE  are more complex and
richer than is normally assumed for "applied mathematics". CT provides a
toolbox of design patterns and principles, whose added value goes beyond
such typical applications of mathematics to SE as formal semantics for a
language, or formal analysis and model checking.

Two aspects of the CT-MDE ``marriage" are discussed in the paper. The
first one is a standard argument about the applicability of a particular
mathematical theory to a particular engineering discipline. To wit, there is
a mathematical framework called CT, there is an engineering domain
called MDE, and we will try to justify the claim that they make a good
match, in the sense that concepts developed in the former are applicable
for mathematical modeling of constructs developed in the latter. What
makes this standard argument exciting is that the mathematical
framework in question is known to be notoriously abstract, while the
engineering domain is very agile and seemingly not suitable for abstract
treatment. Nevertheless, the argument lies within the boundaries of yet
another instance of the evergreen story of applying mathematics to
engineering problems. Below we will refer to this perspective on the issue
as Aspect A.

The second perspective  (Aspect B) is less standard and even more
interesting. It is  essentially based on specific properties of categorical
mathematics and on the observation that software engineering is a special
kind of engineering. To wit, CT is much more than a collection of
mathematical notions and techniques: CT has changed the very way we
build mathematical models and reason about them; it can be seen as a
toolbox of structural design patterns and the guiding principles of their
application. This view on CT is sometimes called \textit{arrow thinking}.
\delok{; it amounts to deploying certain structural modeling skills and
intuitions that could be developed in the course of persistent study of CT,
and the trial and error experience of its applications, rather than being
learned by reading one, or, two, or a dozen papers and books}{not here}
On the other hand, SE, in general, and MDE, in particular, essentially
depend on proper structuring of the universe of discourse
\newok{into subuniverses, which in their turn are further structured and
so on, which finally results in tool architectures and code
modularization}{}. Our experience and attempts to understand complex
structures used in MDE have convinced us that general ideas of arrow
thinking, and general patterns and intuitions of what a healthy structure
should be, turn out to be useful and beneficial for such \del{real,}{}
practical concerns as tool architecture and software design.\new{}{the
pharse is too long. it's be good to shorten}

The paper is structured as follows. In Section~\ref{sect-very-general} we
present two very general A-type arguments\zdmodifyok{, namely}{} that
CT provides a ``right'' mathematical framework for SE. The second
argument also gives strong prerequisites for the B-side of our story.
Section \ref{sect-mde-nutshell} gives a brief outline of MDE,  and Section
\ref{sect-modelware} reveals \zdmodifyok{its truly categorical nature}{a
truly categorical nature of the cornerstone notions of multimodeling and
intermodeling} (another A-argument). In Section
\ref{sect-mde-via-algebra} we present two examples of categorical
arrangement of model management scenarios: model merge and
bidirectional update propagation. This choice is motivated by our research
interests and the possibility to demonstrate the B-side of our story. In
Section \ref{sect-how-ct-used} we discuss and exemplify three ways of
applying CT for MDE: \zdmodifyok{; one is of type A, another is of a mixed
AB nature, and the third one is directly an instance of B. (The taxonomy is
of course somewhat artificial.)}{understanding, design patterns for specific
problems, and general design guidance on the level of tool architecture.}

\section{Two very general perspectives on SE and Mathematics}\label{sect-very-general}

\subsection{The plane of Software $\times$ Mathematics } %

\begin{figure}%
\centering
\includegraphics[width=0.67\columnwidth]{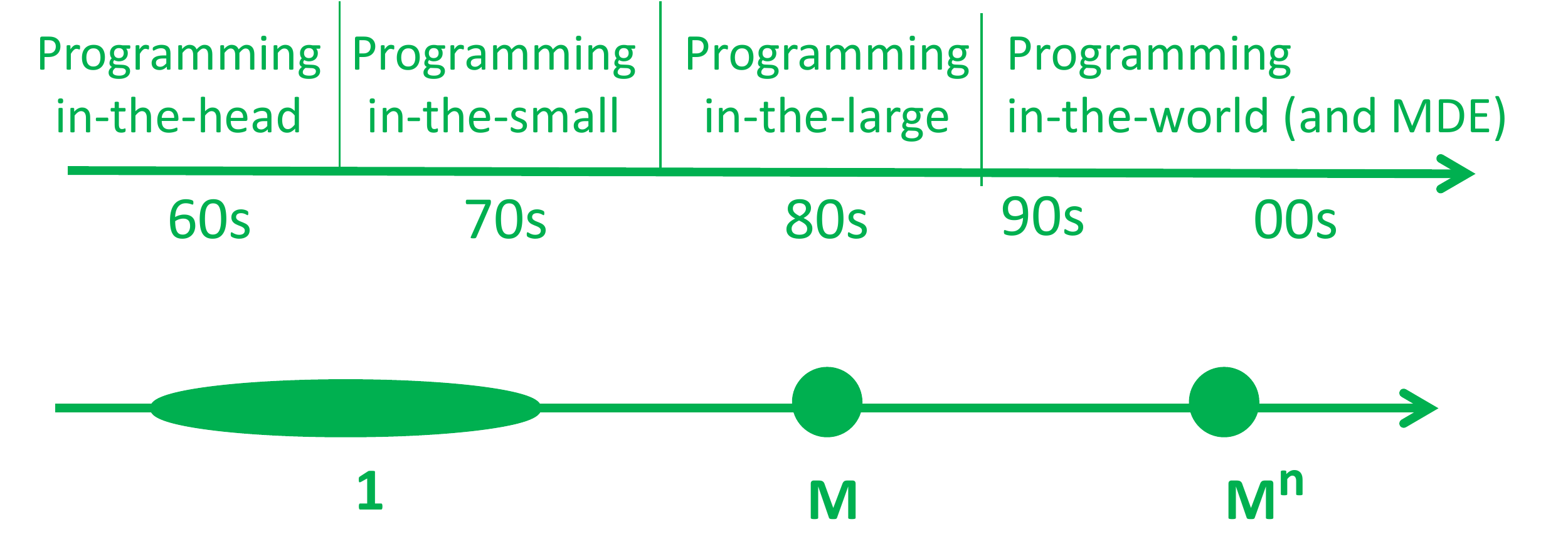}%
\caption{Evolution of software (M  refers to  Many/Multitude)}%
\label{fig:evol-software}%
\end{figure}

The upper half of \figref{fig:evol-software} presents the evolution of
software engineering in a schematic way, following Mary Shaw
\cite{shaw96} and Jos{\'e} Fiadeiro \cite{fiadeiro04}.
Programming-in-the-head refers to the period when a software product
could be completely designed, at least in principle, "inside the head" of one
(super intelligent) programmer, who worked like a researcher rather than
as an engineer. The increasing complexities of problems addressed by
software solutions (larger programs, more complex algorithms and data
structures) engendered more industrially oriented/engineering views and
methods (e.g., structured programming). Nevertheless, for
Programming-in-the-small, the software module remained the primary
goal and challenge of software development, with module interactions
being simple and straightforward (e.g., procedure calls). In contrast,
Programming-in-the-large marks a shift to the stage when module
composition becomes the main issue, with the numbers of modules and the
complexity of module interaction enormously increased. 
%
This tendency continued to grow and widened in scope as time went on,
and today manifests itself as Programming-in-the-world. The latter is
characterized by a large, and growing, heterogeneity of modules to be
composed and methods for their composition, and such essentially
zdmodiffy{in-the-large}{large}{} modern technologies as service
orientation, open source and collaborative software development, and
cloud computing.

The lower part of \figref{fig:evol-software} presents this picture in a very
schematic way as a path from 1 to $M$ to $M^n$ with $M$ referring to
multiplicity in different forms, and degree $n$ indicating the modern
tendencies of growth in heterogeneity and complexity.

{MDE could be seen as a reaction to this \del{historical}{}{}
development, a way of taming the growth of $n$ in a systematic way.
Indeed, until recently, software engineers may feel that they could live
without mathematical models: just build the software by whatever means
available, check and debug it, and keep doing this throughout the
software's life. (Note that the situation in classical (mechanical and
electrical) engineering is essentially different: debugging, say, a bridge,
would be a costly procedure, and classical engineers abandoned this
approach long time ago.) But this gift of easily built systems afforded to
SEs is rapidly degrading as the costs of this process and the liability from
getting it wrong are both growing at an enormous rate. By slightly
rephrasing Dijkstra, we may say that precise modeling and specification
become a matter of
death and life rather than luxury.}{} %

%
\begin{wrapfigure} {r}{0.575\textwidth}%
\vspace{-0.5cm}%
\includegraphics[width=0.575\columnwidth]{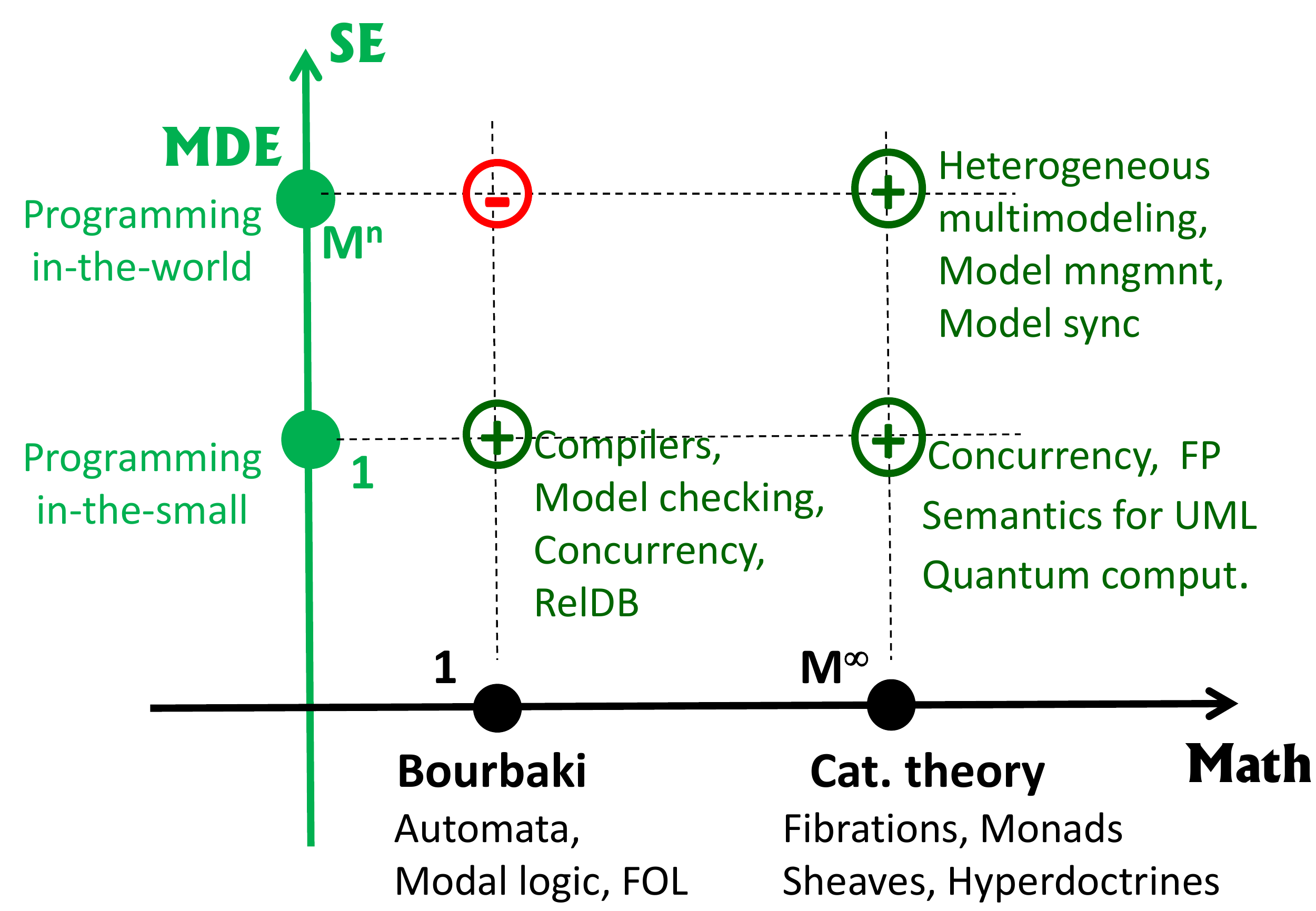}%
\caption{Software engineering and mathematics}%
\label{fig:se-x-math}%
\vspace{-0.2cm}%
\end{wrapfigure}%
%
These considerations give us the vertical axis in \figref{fig:se-x-math},
skipping the intermediate point.  The horizontal axis represents the
evolution of mathematics in a similar simplified way. Point 1 corresponds
to the modern mathematics of mathematical structures in the sense of
Bourbaki: what matters is operations and relations over mathematical
objects rather than their internal structure. Skipped point $M$
corresponds to basic category theory: the internal structure of the entire
mathematical structure is encapsulated, and mathematical studies focus on
operations and relations over structures considered as holistic entities. The
multitude of higher degree, $M^\infty$, refers to categorical facilities for
reflection: enrichment, internalization, higher dimensions, which can be
applied \emph{ad infinitum}, hence, $\infty$-degree.

This (over-simplified) schema gives us four points of Math${\times}$SE
interaction. Interaction (1,1) turned out to be quite successful, as
evidenced by such theory-based practical achievements as compilers,
model checking, and relational DB theory. As for the point $(1,M^n)$,
examining the literature shows that attempts at building theoretical
foundations for MDE based on classical 1-mathematics were not successful.
A major reason seems to be clear: 1-mathematics does not provide an
adequate machinery for specifying and reasoning about inter-structural
relationships and operations, which are at the very heart of modern
software development. This point may also explain the general skepticism
that a modern software engineer, and an academic teaching software
engineering,  feel about the practicality of using mathematics for modern
software design: unfortunately, the only mathematics they know is the
classical mathematics of Bourbaki and Tarski.

On the other hand, we view several recent applications of categorical methods to
MDE problems
\cite{don-models08,boronat08,me-tools08,gabi-models09,%
me-mdi10-springer,rutle-fase10,rossini10,me-jot11,me-fase12,%
rossini12-deepmeta,rutle12} as promising theoretical attempts, with great
potential for practical application. It provides a firm plus for the
$(M^\infty,M^n)$-point in the plane.

Moreover, as emphasized by Lawvere, the strength of CT based modeling
goes beyond modeling multi-structural aspects of the mathematical
universe, and a categorical view of a single mathematical structure can be
quite beneficial too. This makes point $(M^\infty, 1)$ in the plane
potentially interesting, and indeed, several successful applications at this
point are listed in the figure.

\subsection{Mathematical modeling of engineering artifacts:
Round-tripping abstraction vs. waterfall based abstraction}%

\begin{wrapfigure}{r}{0.55\textwidth}%
\vspace{-0.25cm}%
\includegraphics[width=0.5\textwidth]{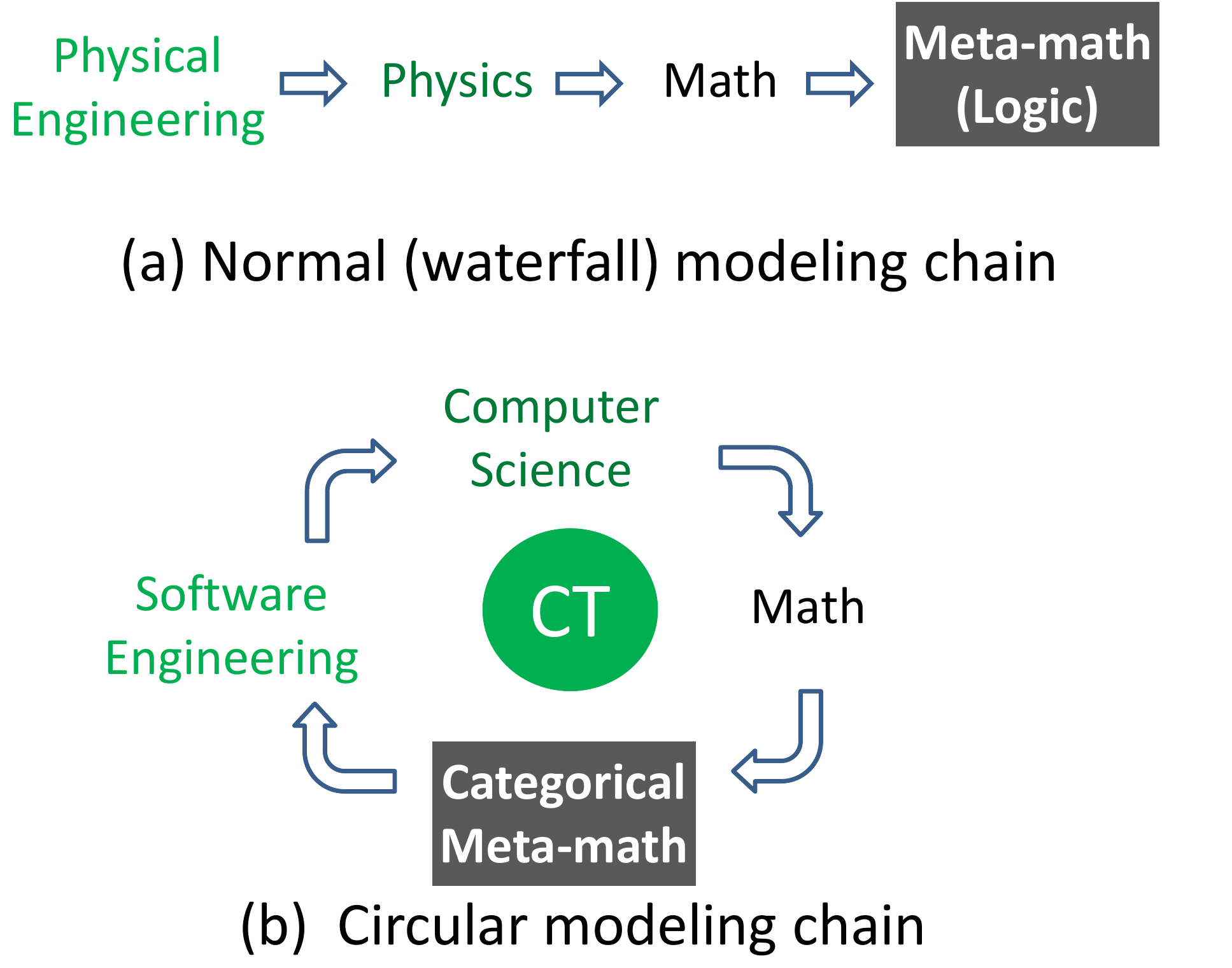}%
\caption{Modeling chains}%
\label{fig:mod-chains}%
\end{wrapfigure}%

Figure \figref{fig:mod-chains}(a) shows a typical way of building
mathematical models for mechanical and electrical engineering domains.
Meta-mathematics (the discipline of modeling mathematical models) is not
practically needed for engineering as such. The situation dramatically
changes for software engineering. Indeed, category theory (CT) could be
defined as a discipline for studying mathematical structures: how to
specify, relate and manipulate them, and how to reason about them. In
this definition, one can safely remove the adjective ``mathematical'' and
consider CT as a mathematical theory of structures in a very broad sense.
Then CT becomes directly applicable to SE as shown in
\figref{fig:mod-chains}(b). Moreover, CT has actually changed the way of
building mathematical structures and thinking about them, and found
extensive and deep applications in theoretical computer science. Hence, CT
can be considered as a common theoretical framework for all modeling
stages in the chain (and be placed at the center). In this way, CT provides
a remarkable unification for modeling activities in SE.

The circular, non linear nature of the figure also illustrates an important
point about the role of CT in SE. Because software artifacts are conceptual
\zdmodiffy{artifacts}{rather than physical entities}{}, there is potential
for feedback between SE and Mathematics in a way that is not possible in
traditional scientific and engineering disciplines. \zdmodiffy{The
mathematics of SE and its discourse, can be, and have been, influenced by
the nature of mathematical formalisms and structures used to talk about
them.}{Design patterns employed in SE can be, and have been, influenced
by mathematical model of software and the way we develop them.}{}

\renewcommand\GG{{\textbf{G}}}
\section{MDE and CT: an overall sketch}

We will begin with a rough general schema of the MDE approach to
building software (Section 3.1), and then will  arrange this schema in
categorical terms (Section 3.2).

\begin{wrapfigure} {r}{0.575\textwidth}%
\vspace{-0.5cm}%
\includegraphics[width=0.5\columnwidth]{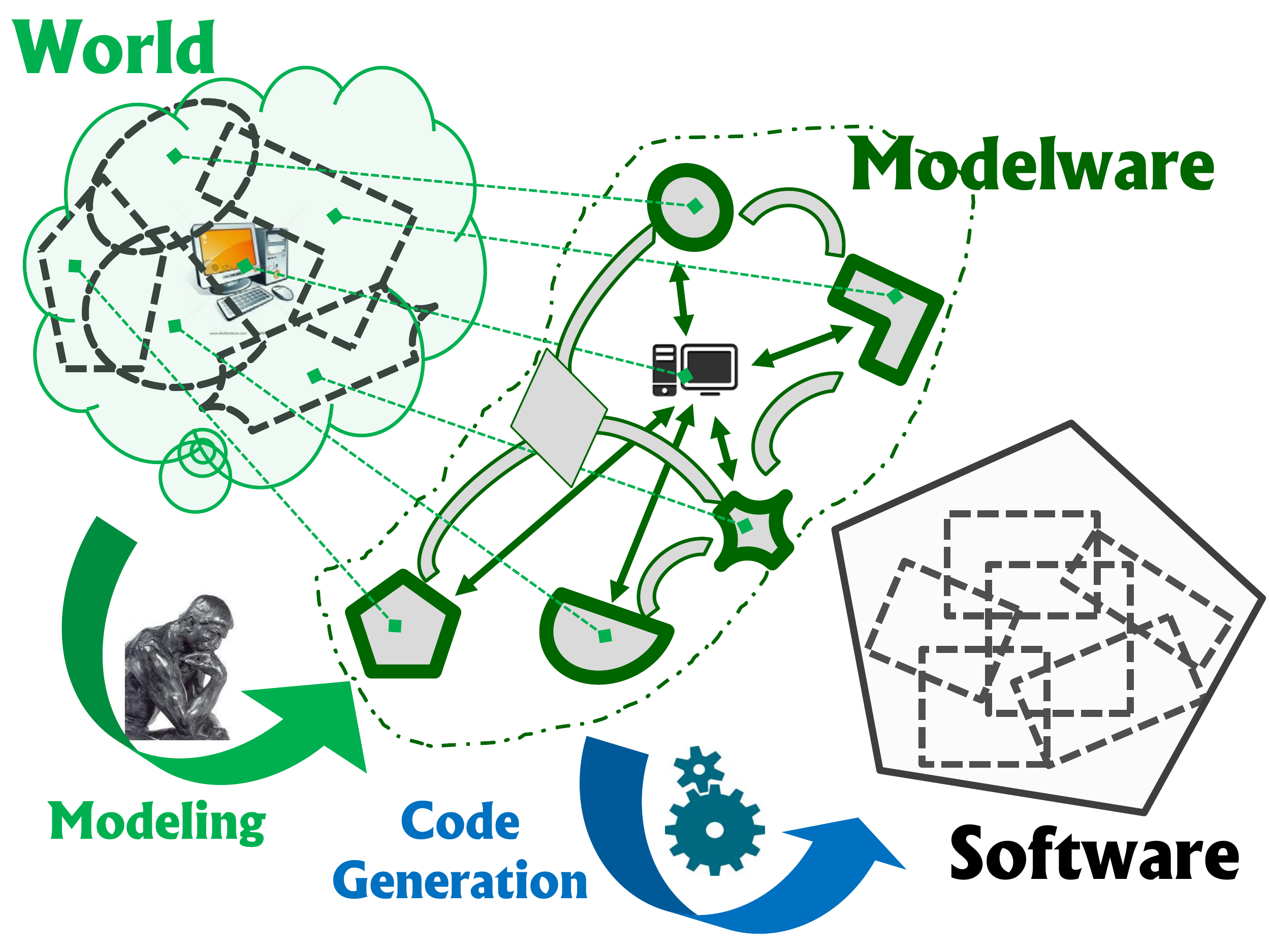}%
\caption{MDE, schematically}%
\label{fig:world2soft}%
\vspace{-0.2cm}%
\end{wrapfigure}%

\subsection{MDE in a nutshell}\label{sect-mde-nutshell}
The upper-left corner of \figref{fig:world2soft} shows  a general goal of
software design: building software that correctly interacts with different
subsystems of the world (shown by figures of different shapes). For
example, software embedded in a car interacts with its mechanical,
electrical and electronic subsystems, with the driver and passengers, and
with other cars on the road in future car designs. These components
interact between themselves, which is schematically shown by overlaps of
the respective shapes. The lower-right corner of \figref{fig:world2soft}
shows software modularized in parallel to the physical world it should
interact with. The passage from the left to the right is highly non-trivial,
and this is what makes SE larger and more challenging than mere
programming. 
An effective means to facilitate the transition is to use models --- a system
of syntactical objects (as a rule, diagrammatic) that serve as abstractions
of the ``world entities'' as shown in the figure (note the links from pieces
of World to the respective parts of
 Modelware). These abstractions are gradually developed and refined until
finally transformed into code. The modelware universe actually consists of
a series of ``modelwares''
---  systems of requirement, analysis, and design models, with  each
consecutive member in the list refining the previous one, and in its own
turn encompassing several internal refinement chains. Modelware
development  consumes intelligence and time, but still easier and more
natural for a human than writing code; the latter is generated
automatically. 
The main idea of MDE is that human intelligence
should be used for building models rather than code.

Of course, models have been used for building software long before the
MDE vision appeared in the market. That time, however, after the first
version of a software product had been released, its maintenance and
further evolution had been conducted mainly through code, so that models
had quickly become outdated, degraded and finally became useless. In
contrast, MDE assumes that maintenance and evolution should also go
through models. No doubts that some changes in the real world are much
easier to incorporate immediately in the code rather than via models, but
then MDE prescribes to update the models to keep them in sync with code.
In fact, code becomes just a specific model, whose only essential
distinction from other models in the modelware universe is its final
position in the refinement chain. Thus, the Modelware boundary in
\figref{fig:world2soft} should be extended to encompass the Software
region too.

\subsection{Modelware categorically}\label{sect-modelware}

Consider a modelware snapshot in \figref{fig:world2soft}. Notice that
models as such are separated whereas their referents are overlapped, that
is, interact between themselves. This interaction is a fundamental feature
of the real world, and to make the model universe adequate to the world,
intermodel correspondences/relations must be precisely specified. (For
example, the figure shows three binary relations, and one ternary relation
visualized as a ternary span with a diamond head.) With reasonable
modeling techniques, intermodel relations should be compatible with
model structures. The modelware universe then appears as a collection of
structured objects and structure-compatible mappings between them, that
is, as a categorical phenomenon. In more detail, a rough categorical
arrangement could be as follows.

\mysubsubsection{The base universe.} Models are multi-sorted structures
whose theories are called \emph{metamodels}. The latter can be seen as
generalized sketches \cite{Makkai97,me-entcs08}, that is, pairs
$M=(G_M,C_M)$ with $G_M$ a graph (or, more generally, an object of an
apiori fixed presheaf topos \GG), and $C_M$ a set of \emph{constraints}
(\ie, diagram predicates) declared over $G_M$. An \emph{instance} of
metamodel $M$ is a pair $A=(G_A,t_A)$ with $G_A$ another graph (an
object in \GG) and \flar{t_A}{G_A}{G_M} a mapping (arrow in \GG) to
be thought of as \emph{typing}, which satisfy the constraints,  $A\models
C_M$ (see \cite{me-entcs08} for details). An \emph{instance mapping}
$A\rightarrow B$ is a graph mapping  \flar{f}{G_A}{G_B} commuting
with typing:
$f;t_B=t_A$. 
This defines a category $\modcat(M)\subset\GG/G_M$ of $M$-instances.

To deal with the \hgs\  situation of models over different metamodels, we
first introduce metamodel morphisms \flar{m}{M}{N} as sketch
morphisms, \ie, graph mappings \flar{m}{G_M}{G_N} compatible with
constraints. This gives us a category of metamodels \mmodcat. Now we
can merge all categories $\modcat(M)$ into one category \modcat, whose
objects are instances (= $\GG$-arrows) \flar{t_A}{G_A}{G_{M(A)}},
\flar{t_B}{G_B}{G_{M(B)}} \etc, each having its metamodel, and
morphisms \flar{f}{A}{B}  
are pairs \flar{f_{\mathrm{data}}}{G_A}{G_B},
\flar{f_{\mathrm{meta}}}{M(A)}{M(B)} such that
$f_{\mathrm{data}};t_B=t_A;f_{\mathrm{meta}}$, \ie, commutative
squares in \GG. Thus, $\modcat$ is a subcategory of the arrow category
$\GG^{\cdot\rightarrow\cdot}$.

It can be shown that pulling back a legal instance \flar{t_B}{G_B}{G_N}
of metamodel $N$ along a sketch morphism \flar{m}{M}{N} results in a
legal instance of $M$ \cite{me-entcs08}.
We thus have a fibration 
\flar{\projfun}{\modcat}{\mmodcat}, 
whose Cartesian lifting is given by pullbacks.

\mysubsubsection{Intermodel relations and queries.} A typical
intermodeling situation is when an element of one model corresponds to
an element that is not immediately present in another model, but can be
derived from other elements of that model by a suitable operation (a
query, in the database jargon) \cite{me-fase12}.
Query facilities can be modeled by a pair of monads $(\qmond,\qmon)$
over categories \mmodcat\ and \modcat, resp. The first monad describes
the syntax (query definitions), and the second one provides the semantics
(query execution). 

A fundamental property of queries is that the original data are not
affected: queries compute new data but do not change the original.
Mathematical modeling of this property results in a number of equations,
which can be summarized by saying that monad \qmon\  is
\projfun-Cartesian, \i.e., the Cartesian and the monad structure work in
sync. 
If can be shown \cite{me-fase12} that a query language
$(\qmon,\qmond)$ gives rise to a fibration
\flar{\pfiq{}}{\modcat_{\qmon}}{\mmodcat_{\qmond}} between the
corresponding Kleisli categories. These Kleisli categories have immediate
practical interpretations. Morphisms in $\mmodcat_{\qmond}$ are
nothing but view definitions: they map elements of the source metamodel
to queries against the target one. Correspondingly, morphisms in
$\modcat_{\qmon}$ are view executions composed from query execution
followed by retyping. The fact that projection \pfiq{} is fibration implies
that the view execution mechanism is compositional: execution of a
composed view equals the composition of executions.

\begin{wrapfigure}{R}{.575\textwidth}
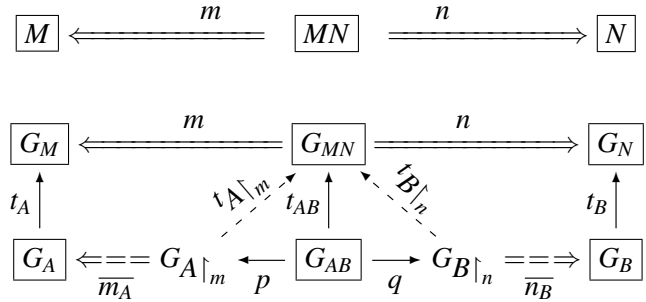

\vspace{-0.5ex}%
\centering%
\begin{tabular}{c}
\begin{diagram}[small]
\dbox{M} &&&\lViewto^{m} && ~~~\dbox{MN}~~~ &&\rViewto^{n}&&&\dbox{N}\\%
\end{diagram}   %
\\[20pt]
 \begin{diagram}[small]
\dbox{G_M} &&\lViewto^{m} && \dbox{G_{MN}} &&\rViewto^{n}&&\dbox{G_N}\\%
\uTo<{\tymap{A}} &&&\ruDashto^{t_{\cartliob{A}{m}}} &\uTo<{\tymap{AB}}
&\luDashto^{t_{\cartliob{B}{n}}} &&&\uTo<{\tymap{B}} \\%
\dbox{G_A}&\lDerviewto_{\cartliarr{m}{A}} & G_{\cartliob{A}{m}} &\lTo_p %
& \dbox{G_{AB}}
  &\rTo_q &G_{\cartliob{B}{n}} &\rDerviewto_{\cartliarr{n}{B}} & \dbox{G_B}\\%
\end{diagram}%
\end{tabular}  %
\caption{Correspondences between \hgs\ models%
\label{fig:hgs-matching}}%
\end{wrapfigure}%

Now a correspondence between models $A,B$ over metamodels $M,N$ can
be specified by data shown in \figref{fig:hgs-matching}; these data consist
of three components..
(1) span $(\bdlar{m}{MN}{N},\,\dlar{n}{MN}{N})$ (whose legs are
    Kleisli  mappings)  specifies a  common view $MN$ between the two metamodels.  %
(2) trapezoids (arrows in $\modcat_{\qmon}$) are produced by
    \pfiq{}-Cartesian ``lifting'', \ie, by executing views $m$ and $n$
    for models $A$ and $B$ resp., which results in models %
    \cartliob{A}{m} and
    \cartliob{B}{n} (here and below we use the
    following notation: computed nodes are  not framed, and computed
    arrows are dashed).  %
(3) span $(\bflar{p}{AB}{\cartliob{A}{m}},\,
\flar{q}{AB}{\cartliob{B}{n}})$ specifies a correspondence between the
views. Note that this span is an independent modelware component and cannot
be derived from models $A, B$.

Spans like in \figref{fig:hgs-matching} integrate a collection of models
into a holistic system, which we will refer to as a \emph{multimodel}.
Examples, details, and a precise definition of a multimodel's consistency
can be found in \cite{me-mdi10-springer}. 


It is tempting to encapsulate spans in \figref{fig:hgs-matching} as
composable arrows and work with the corresponding (bi)categories of
metamodels and models. Unfortunately, it would not work out because, in
general, Kleisli categories are not closed under pullbacks, and it is not
clear how to compose Kleisli spans. It is an important problem to overcome
this obstacle and find a workable approach to Kleisli spans,

Until the problem above is solved, our working universe is the Kleisli
category of \hgs\ models fibred over the Kleisli category of metamodels.
This universe is a carrier of different operations and predicates over
models, and a stage on which different modeling scenarios are played.
Classification and specification of these operations and predicates, and
their understanding in conventional mathematical terms, is a major task of
building mathematical foundations for MDE. Algebraic patterns appear
here quite naturally, and then model management scenarios can be seen
as algebraic terms composed from diagram-algebra operations over models
and model mappings.\footnote{Note, however, that a proper categorical
treatment of these operations in terms of universal constructions can be
not straightforward.} The next section provides examples of such algebraic
arrangements.

\section{Model management (MMt) and algebra: Two examples}\label{sect-mde-via-algebra}


We will consider two examples of algebraic modeling of MMt scenarios. A
simple one --- model merging, and a more complex and challenging ---
bidirectional update propagation (BX).

\subsection{Model merge via colimit}\label{sect-model-merge}

\begin{wrapfigure} {r}{0.326\textwidth}%
\includegraphics[width=0.325\columnwidth]{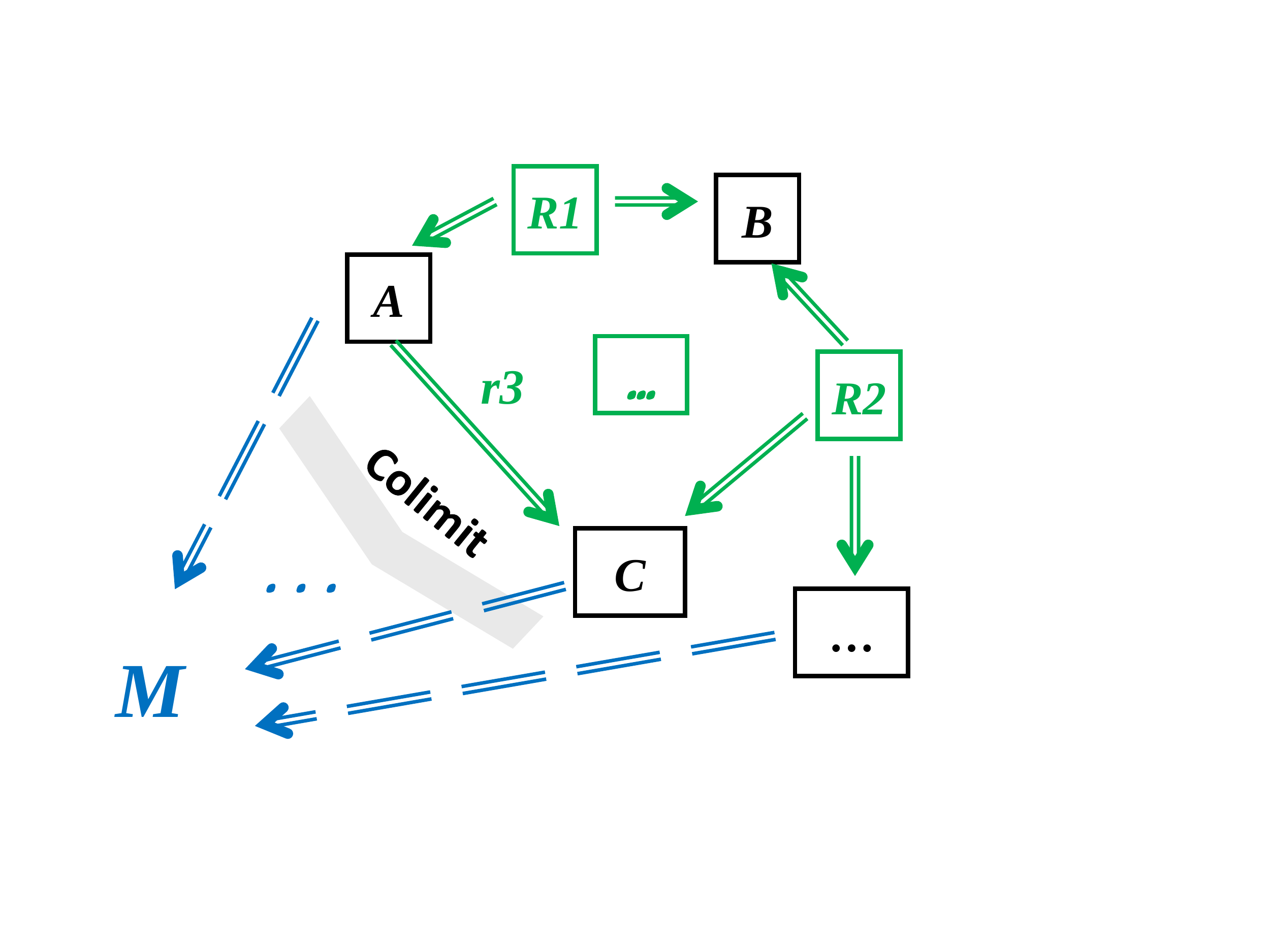}%
\caption{Model merge}%
\label{fig:model-merge}%
\vspace{-0.2cm}%
\end{wrapfigure}

Merging several interrelated models without data redundancy and loss is
an important MDE scenario. Models are merged (virtually rather than
physically)  to check their consistency, or to extract an integrated
information about the system. A general schema is shown in
\figref{fig:model-merge}. Consider first the case of several homogeneous
models $A,B,C...$ to be merged. The first step is to specify
correspondences/relations between models via Kleisli spans $R1, R2,...$,
or perhaps direct mappings like $r3$. The intuition of merging without
data loss and redundancy (duplication of correspondent data) is precisely
captured by the universal property of colimits, that is, it is reasonable to
define merge as the colimit of a diagram of models and model mappings
specifying intermodel correspondences.

If models are \hgs, their relations are specified as in
\figref{fig:hgs-matching}. To merge, we first merge metamodels modulo
metamodel spans. Then we can consider all models and heads of the
correspondence spans as instances of the merged metamodel, and merge
models by taking the colimit of the entire diagram in the category of
instances of the merged metamodel.

An important feature of viewing model merge as described above is a clear
separation of two stages of the merge process: (i) discovery and specifying intermodel
correspondences (often called model matching), and (ii) merging models
modulo these correspondences. The first stage is inherently heuristic and
context dependent. It can be assisted by tools based on AI-technologies,
but in general a user input is required for final adjustment of the match (and of course to define the heuristics used by the tool).
The second stage is pure algebra (colimit) and can be performed automatically.
 The first step may heavily depend on the domain  and the application,
 while the second one is domain and application independent. However,
 a majority of model merge tools combine the two stages into a
holistic merge algorithm, which first somehow relates models based on a
specification of conflicts between them, and then proceeds accordingly to
merging. Such an approach complicates merge algorithms, and makes a
taxonomy of conflicts their crucial component; typical examples are
\cite{spacca94,pottinger03}.

\new{The cause of this deficiency is that tool builders rely on a very simple
notion of model matching, which amounts to linking
\emph{the-same-semantics} elements in the models to be matched.
However,  as discussed above in Section \ref{sect-modelware}, for an
element $e$ in model $A$, the-same-semantics $B$-element $e'$ can
only be indirectly present in $B$, \ie, $e'$ can be derived from other
elements of $B$ with a suitable operation (query) over $B$ rather than
being an immediate element of $B$. With complex (Kleisli) matching that
allows one to link basic elements in one model with derived elements in
another model, the algebraic nature of merge as such (via the colimit
operation) can be restored. Indeed, it is shown in \cite{myTR-integr08}
that all conflicts considered in \cite{pottinger03} can be managed via
complex matching, that is,  described via
Kleisli spans with a suitable choice of queries, afterwards merge is computed via
colimit.  
}%
%

\subsection{Bidirectional update propagation (BX)}\label{sect-bx}

Keeping a system of models mutually consistent ({model \syncn}) is vital
for model-driven engineering. In a typical scenario, given a pair of
inter-related models, changes in either of them are to be propagated to
the other to restore consistency. This setting is often referred to as
bidirectional model transformation (BX)
\cite{
Czarnecki09a}.

\subsubsection{BX via tile algebra}
\begin{wrapfigure} {r}{0.5\textwidth}%
\includegraphics[width=0.475\columnwidth]{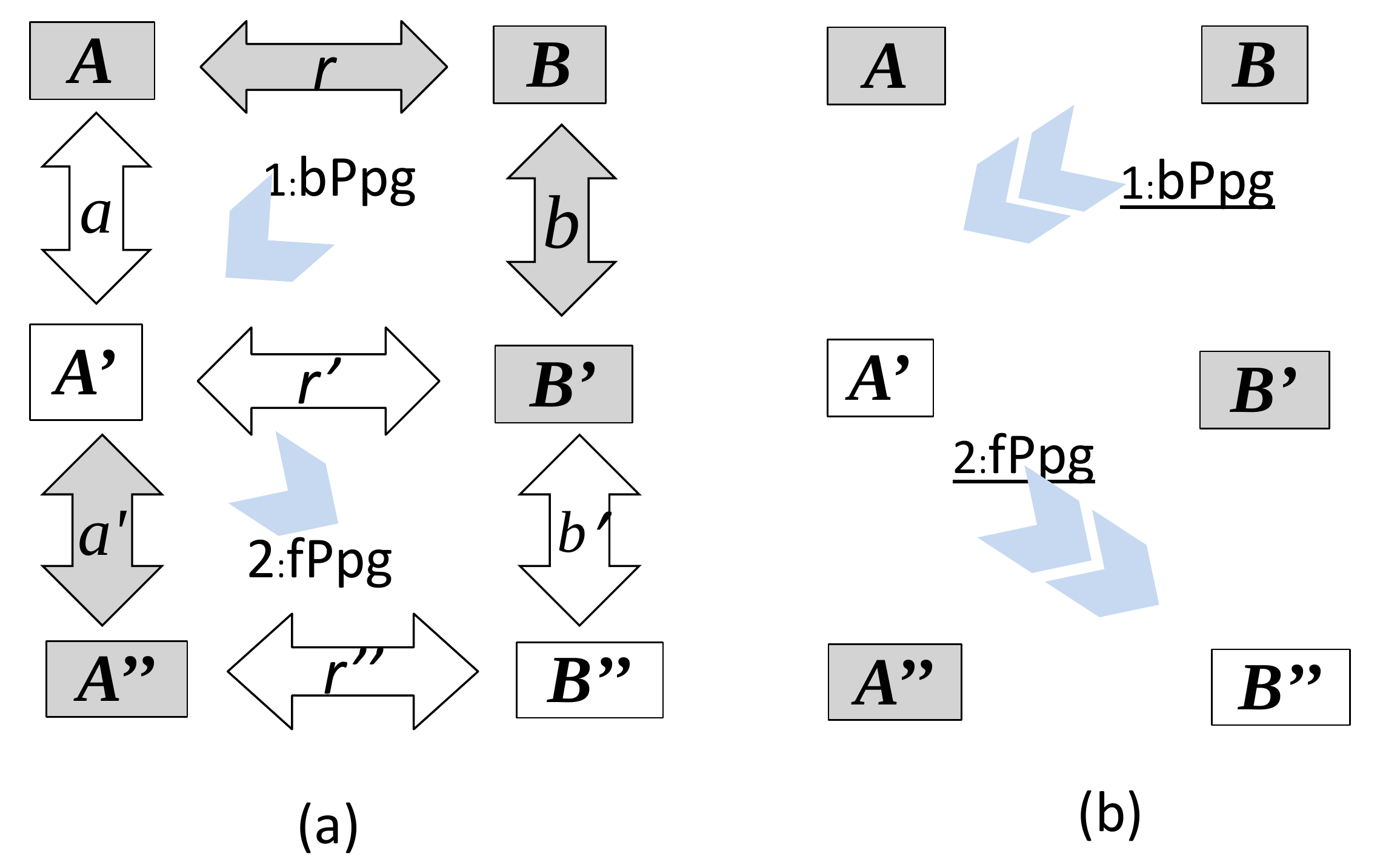}%
\caption{BX scenario specified in (a) delta-based and (b) state-based way}%
\label{fig:bx-scenario}%
\vspace{-0.2cm}%
\end{wrapfigure}

A simple BX-scenario is presented in \figref{fig:bx-scenario}(a). Two
models, $A$ and $B$, are interrelated by some \emph{correspondence
specification} 
$r$ (think of a span in a suitable
category, or an object in a suitable comma category, see
\cite{me-mdi10-springer} for examples). We will often refer to them as
\emph{horizontal deltas} between models. In addition, there is a notion of
delta  \emph{consistency} (extensionally, a class of \emph{consistent}
deltas), and if $r$  is consistent, we call models $A$ and $B$
synchronized.

Now suppose that (the state of) model $B$ has changed: the updated
(state of the) model is $B'$, and arrow $b$ denotes the correspondence
between $B$ and $B'$ (\emph{a vertical delta}). The reader may think of
a span, whose head consists of unchanged elements  and the legs are
injections so that $B$'s elements beyond the range of the upper leg are
deleted, and $B'$'s elements beyond the range of the lower leg are inserted.
Although update spans are denoted by bidirectional arrows, the upper
node is always the source, and the lower is the target.

Suppose that we can re-align models $A$ and $B'$ and compute new
horizontal delta $r*b$ (think of a span composition). If this new delta is
not consistent, we need to update model $A$ so that the updated model
$A'$ would be in sync with $B'$. More accurately, we are looking for an
update \bilar{a}{A}{A'} such that the triple $(A', r', B')$ is consistent. Of
course, we want to find a minimal update $a$ (with the biggest head) that
does the job.

Unfortunately, in a majority of practically interesting situations, the
minimality condition is not strong enough to provide uniqueness of $a$. To
achieve uniqueness, some update propagation policy is to be chosen, and
then we have an algebraic operation \bppg\ ('b' stands for 'backward'),
which, from a given a pair of arrows $(b,r)$ connected as shown in the
figure, computes
another pair $(a,r')$ connected with $(b,r)$ as shown in the figure. 
Thus, a propagation policy {is} algebraically modeled by a diagram
operation of arity specified by the upper square in
\figref{fig:bx-scenario}(a): shaded elements denote the input data,
whereas blank ones are the output. Analogously, choosing a forward
update propagation policy (from the $A$-side to the $B$-side) provides a
forward operation \fppg\ as shown by the lower square.

The entire scenario is a composition of two operations: a part of the input
for \new{operation application}{ZD} 2:\fppg\ is provided by the output of
1:\bppg. In general, composition of diagram operations, \ie, operations
acting upon configurations of arrows (diagrams), amounts to their
\emph{tiling}, as shown in the figure; then complex \syncon\ scenarios
become \emph{tiled} structures. Details, precise definitions and examples
can be found in \cite{me-gttse-long}. 


Different diagram operations involved in model \syncon\ are not
independent and their interaction must satisfy certain conditions. These
conditions capture the semantics of \syncon\ procedures, and their
understanding is important for the user of \syncon\ tools: it helps to avoid
surprises when automatic \syncon\ steps in. Fortunately, principal
conditions (\syncon\ laws) can be formulated as universally valid
equations between diagrammatic terms --- a tile algebra counterpart of
universal algebraic \emph{identities}. In this way BX becomes based on
an algebraic theory: a signature of diagram operations and a number of
equational laws they must satisfy. The appendix presents one such theory
--- the notion of a \emph{symmetric delta lens}, which is currently an
area of active research from both a practical and a theoretical perspective.

\subsubsection{BX: delta-based vs. state-based}

As mentioned above, understanding the semantics of model \syncon\
procedures is important, both theoretically and practically. Synchronization
tools are normally built on some underlying algebraic theory
\cite{Foster07,Xiong07,Matsuda07,Foster08,gsdlab-tse09,%
QVT08,frank-models11%
}, and many such tools (the first five amongst those cited above) use
algebraic theories based on state-based rather than delta-based
operations. The state-based version of the propagation scenario in
\figref{fig:bx-scenario}(a) is described in \figref{fig:bx-scenario}(b). The
backward propagation operation takes models $A,B,B'$, computes
necessary relations between them ($r$ and $b$ on the adjacent diagram),
and then computes an updated model $A'$. The two-chevron symbol
reminds us that the operation actually consists of two stages: model
alignment (computing $r$ and $b$) and update propagation as such.

The state-based \fwk s, although they may look simpler, actually hides several
serious deficiencies. Model alignment is a difficult task that requires
contextual information about models.  It can be facilitated by intelligent
AI-based tools, or even be automated, but the user should have an option
to step in and administer corrections. In this sense, model alignment is similar to model
matching preceding model merge.%
\footnote{A difference is that model matching usually refers to relating
independently developed models, while models to be aligned are
often connected by a given transformation.} 
Weaving alignment (delta discovery) into update (delta) propagation
essentially complicates the semantics of the latter, and correspondingly
complicates the algebraic theory. In addition, the user does not have an
access to alignment results and cannot correct them.

\begin{wrapfigure} {R}{0.326\textwidth}%
\vspace{-0.5cm}%
\includegraphics[width=0.325\columnwidth]{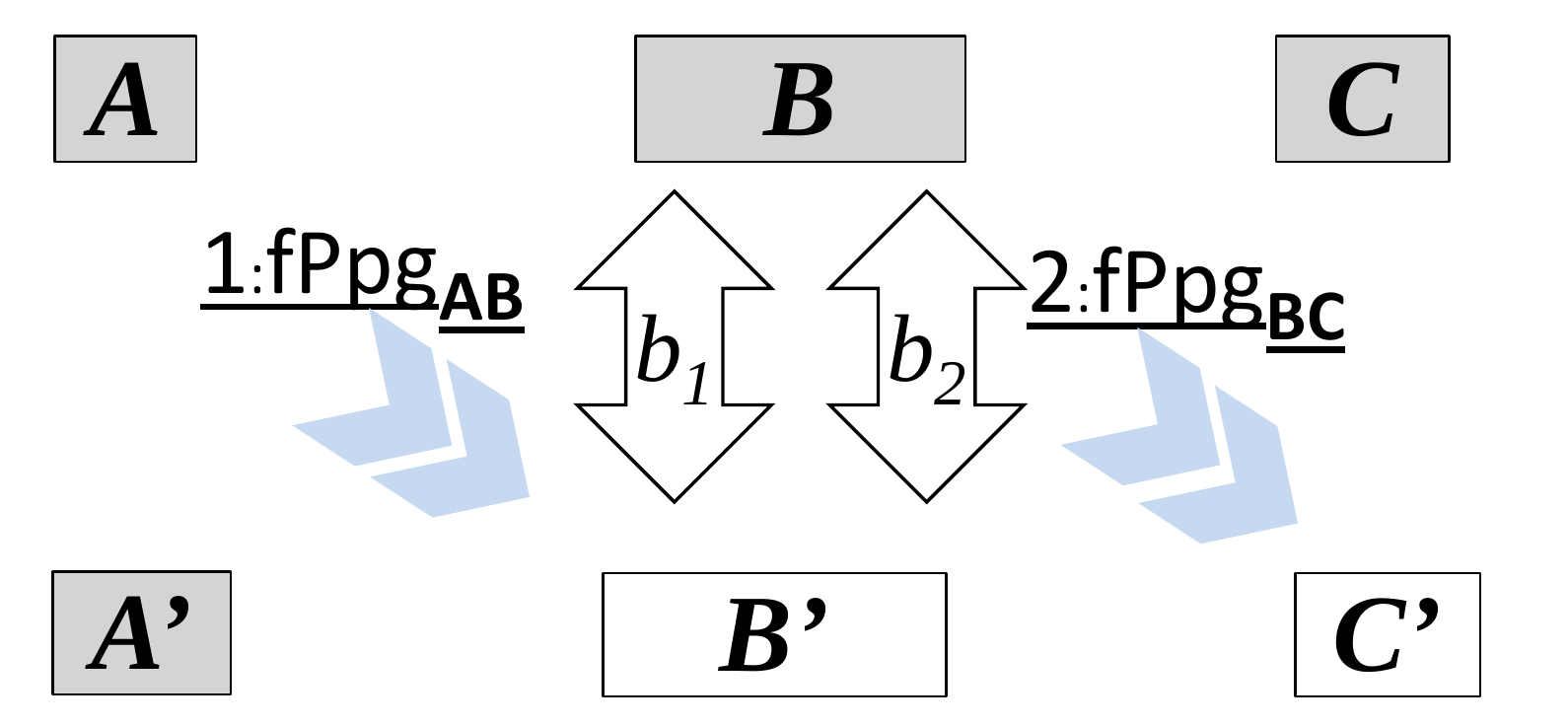}%
\caption{State-based BX: erroneous horizontal composition}%
\label{fig:bx-sttFwk-errors}%
\vspace{-0.2cm}%
\end{wrapfigure}%

Two other serious problems of the state-based \fwk s and architectures are
related to operation composition. The scenario described in
\figref{fig:bx-scenario}(a) assumes that the model correspondence (delta)
used for update propagation 2:\fppg\ is the delta computed by operation
1:\bppg; this is explicitly specified in the tile algebra specification of the
scenario. In contrast, the state-based \fwk\ cannot capture this
requirement. A similar problem appears when we sequentially compose a
BX program synchronizing models A and B and another program
synchronizing models B and C: composition amounts to horizontal
composition of propagation operations as shown in
\figref{fig:bx-sttFwk-errors}, and again continuity, $b_1=b_2$,  cannot
be specified in the state-based \fwk. A detailed discussion of delta- vs.
state-based \syncon\ can be found in \cite{me-jot11,me-models11}.

\subsubsection{Assembling model transformations}

\begin{wrapfigure}{R}{.375\textwidth}
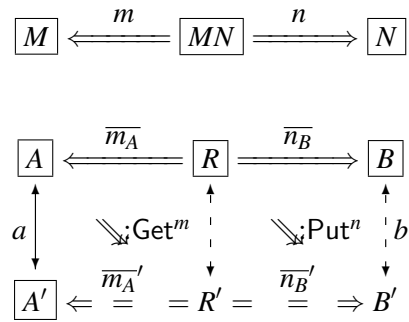

\vspace{-0.5cm}%
\begin{tabular}{c}
\begin{diagram}[w=1.1\cellW,h=1.1\cellH] 
\dbox{M} &\lViewto^{m} & \dbox{MN} &\rViewto^{n}&\dbox{N}\\%
\end{diagram}   %
\\[25pt]
\begin{diagram}[w=1.1\cellW,h=1.1\cellH] 
\dbox{A} & \lViewto^{\cartliarr{m}{A}}  & \dbox{R} %
               & \rViewto^{\cartliarr{n}{B}} & \dbox{B}%
            \\%
\dUpdto<a & \seTilearrow{\get{m}}~~~ & \dDerupdto>{} %
                 & \seTilearrow{\put{n}}~~~ & \dDerupdto>{b}
             \\%
\dbox{A'} & \lDerviewto^{\cartliarr{m}{A}'}  & R' %
               & \rDerviewto^{\cartliarr{n}{B}'}   & B' %
\end{diagram}%
\end{tabular}  %
\caption{Model transformation via GetPut-decomposition%
\label{fig:trafo-as-getput}}%
\end{wrapfigure}

Suppose $M,N$ are two metamodels, and we need to transform
$M$-instances (models) into $N$-ones. Such a transformation makes
sense if metamodels are somehow related, and we suppose that their
relationship is specified by a span $(\bdlar{m}{MN}{M},\;
\dlar{n}{MN}{N})$  (\figref{fig:trafo-as-getput}), whose legs are Kleisli
mappings of the respective query monad. %
\newarrow{Viewto}===={=>}
\newarrow{Derviewto}{=}{}{=}{}{=>}
\newarrow{Updto}{<}---{>}
\newarrow{Derupdto}{<}{dash}{}{dash}{>}

Now $N$-translation of an $M$-model $A$ can be done in two steps. First,
view $m$ is executed (via its Cartesian lifting actually going down in the
figure), and we obtain Kleilsi arrow \bdlar{\cartliarr{m}{A}}{R}{A}
(with $R=\cartliob{A}{m}$). 
Next we need to find an $N$-model $B$ such that its view along $n$,
\cartliob{B}{n}, is equal to $R$. In other words, given a view, we are
looking for a source providing this view. There are many such sources, and
to achieve uniqueness, we need to choose some policy. Afterwards, we
compute model $B$ related to $A$ by span $(\cartliarr{m}{A},
\cartliarr{n}{B})$.

If model $A$ is updated to $A'$, it is reasonable to compute a
corresponding update \bilar{b}{B}{B'} rather than recompute $B'$ from
scratch (recall that models can contain thousands elements). Computing
$b$ again consists of two steps shown in the figure. Operations \get{m}
and \put{n} are similar to \fppg\ and \bppg\ considered above, but work
in the asymmetric situation when mappings $m$ and $n$ are total (Kleisli)
functions and hence view $R$ contains nothing new wrt. $M$ and $N$.
Because of asymmetry, operations \get{} (`get' the view update) and
\put{} ('put' it back to the source) are different. \get{m} is uniquely
determined by the view definition $m$. \put{n} needs, in addition to $n$,
some update propagation policy. After the latter is chosen, we can realize
transformation from $M$ to $N$  incrementally by composition
$\fppg=\get{m};\put{n}$ ---  this is an imprecise linear notation for tiling
(composition of diagram operations) specified in
\figref{fig:trafo-as-getput}. 

Note that the initial transformation from $M$ to $N$ sending, first, an
$M$-instance  $A$ to its view $R=\cartliob{A}{m}$, and then finding an
$N$-instance $B\in N$ such that $\cartliob{B}{n}=R$, can be also
captured by \get{} and \put{}. For this, we need to postulate initial
objects  $\Omega_M$ and $\Omega_N$ in categories of $M$- and
$N$-instances, so that for any $A$ over $M$ and $B$ over $N$ there are
unique updates  \flar{0_A}{\Omega_M}{A} and
\flar{0_B}{\Omega_N}{B}. Moreover, there is a unique span
$(\bdlar{m_\Omega}{\Omega_{MN}}{\Omega_M},\;
\dlar{n_\Omega}{\Omega_{MN}}{\Omega_N})$ relating these initial
objects. Now, given a model $A$, model  $B$ can be computed as $B'$ in
\figref{fig:trafo-as-getput} with the upper span being $(m_\Omega,
n_\Omega)$, and the update $a$ being \flar{0_A}{\Omega_M}{A}.

The backward transformation is defined similarly by swapping the roles of
$m$ and $n$: $$\bppg=\get{n};\put{m}.$$

The schema described above can be seen as a general pattern for defining
model transformation declaratively with all benefits (and all pains) of
having a precise specification before the implementation is approached
(and must obey). Moreover, this schema can provide some semantic
guarantees in the following way. Within the tile algebra \fwk, laws for
operations \get{} and \put{}, and their interaction (invertibility), can be
precisely specified \cite{me-jot11} (see also the discussion in Section
5.1); algebras of this theory are called \emph{delta lenses}. Then we can
deduce the laws for the composed operations \fppg\ and \bppg\ from the
delta lens laws. Also, operations \get{m}, \put{m} can themselves be
composed from smaller blocks, if the view $m$ is composed:
$m=m_1;m_2;...;m_k$, via sequential lens composition. In this way, a
complex model transformation is assembled from elementary
transformation blocks, and its important semantic properties are
guaranteed. More examples and details can be found in
\cite{me-gttse-long}.

\renewcommand\get[1]{\mbox{\diagopername{Get}$^{#1}$}}
\renewcommand\put[1]{\mbox{\diagopername{Put}$^{#1}$}}

\section{Applying CT to MDE: Examples and Discussion.}\label{sect-how-ct-used}

We will try to exemplify and discuss three ways in which  CT can be
applied in MDE. The first one --- gaining a deeper understanding of an
engineering problem --- is standard, and appears as a particular
instantiation of the general case of CT's employment in applied domains.
The other two are specific to SE: structural patterns provided by
categorical models of the software system to be built can directly influence
the design. \new{We will use models of BX as our main benchmark; other
examples will be also used when appropriate.}{}

\mysubsubsection{5.1  Deeper understanding.} As mentioned in
\sectref{sect-bx}, stating algebraic laws that BX procedures must obey is
practically important as it provides semantic guaranties for \syncon\
procedures. Moreover, formulation of these laws should be semantically
transparent and concise as the user of \syncon\ tools needs a clear
understanding of propagation semantics. The original state-based theory of
asymmetric BX \cite{Foster07} considered two groups of laws: invertibility
(or round-tripping) laws, \getputlaw\ and \putgetlaw,  and history
ignorance, \putputlaw . Two former laws say that two propagation
operations, \get{}
and \put{},  
are mutually inverse. The \putputlaw\ law says that if a complex update is
decomposed into consecutive pieces, it can be propagated incrementally,
one piece after the other. A two-sorted algebra comprising two operations,
\get{} and \put{}, satisfying the laws, is called a \emph{well-behaved
lens}. 

Even an immediate arrow-based generalization of lenses to delta lenses
(treated in elementary terms via tile algebra \cite{me-gttse-long,%
me-jot11}) revealed that the \getputlaw\ law is a simple law of identity
propagation, \idputlaw, rather than of round-tripping. The benefits of
renaming \getputlaw\ as \idputlaw\ are not exhausted by clarification of
semantics: as soon as we understand that the original \getputlaw\ is about
identity propagation, we at once ask what the \emph{real round-tripping}
law \getputlaw\ should be, and at once see that operation \put{} is not
the inverse of \get{}. We only have the weaker 1.5-round-tripping
\getputgetlaw\ law (or \emph{weak} invertibility; see the Appendix,
where the laws in question are named \idppglaw\ and \fbfppglaw\ and
\bfbppglaw). It is interesting (and remarkable) that papers
\cite{{me-models08},Hofmann11}, in which symmetric lenses are studied
in the state-based setting, mistakenly consider identity propagation laws
as round-tripping laws, and correspondingly analyze a rather poor
BX-structure without real round-tripping laws at all.

The tile algebra formulation of the \putputlaw\ law clarified its meaning as
a composition preservation law \cite{me-gttse-long,me-jot11}, but did not
solve the enigmatic \putputlaw\  problem. The point is that \putputlaw\
does not hold in numerous practically interesting situations, but its entire
removal from the list of BX laws is also not satisfactory, as it leaves
propagation procedures without any constraints on their compositionality.
The problem was solved, or at least essentially advanced, by a truly
categorical analysis performed by Michael Johnson et al
\cite{joro10,joro12}. They have shown that an asymmetric well-behaved
lens is an algebra for some KZ monad, and \putputlaw\ is nothing but the
basic associativity condition for this algebra. Hence, as Johnson and
Rosebrugh write in \cite{joro12}, the status of the \putputlaw\ changes
from being (a)  ``some law that may have arisen from some special
applications and should be discarded immediately if it seems not to apply
in a new application'' to (b) a basic requirement of an otherwise adequate
and general mathematical model. And indeed, Johnson and Rosebrugh
have found a weaker --- \emph{monotonic} ---  version of \putputlaw\
(see \figref{fig:monot-PpgPpg} in the Appendix), which holds in a majority
of practical applications, including those where the original
(non-monotonic or mixed) \putputlaw\ fails. Hopefully, this categorical
analysis can be generalized for the symmetric lens case, thus stating solid
mathematical foundations for BX.

\mysubsubsection{5.2  Design patterns for specific problems.} \medskip

\begin{wrapfigure} {R}{0.405\textwidth}%
\vspace{-0.5cm}%
\includegraphics[width=0.4\columnwidth]{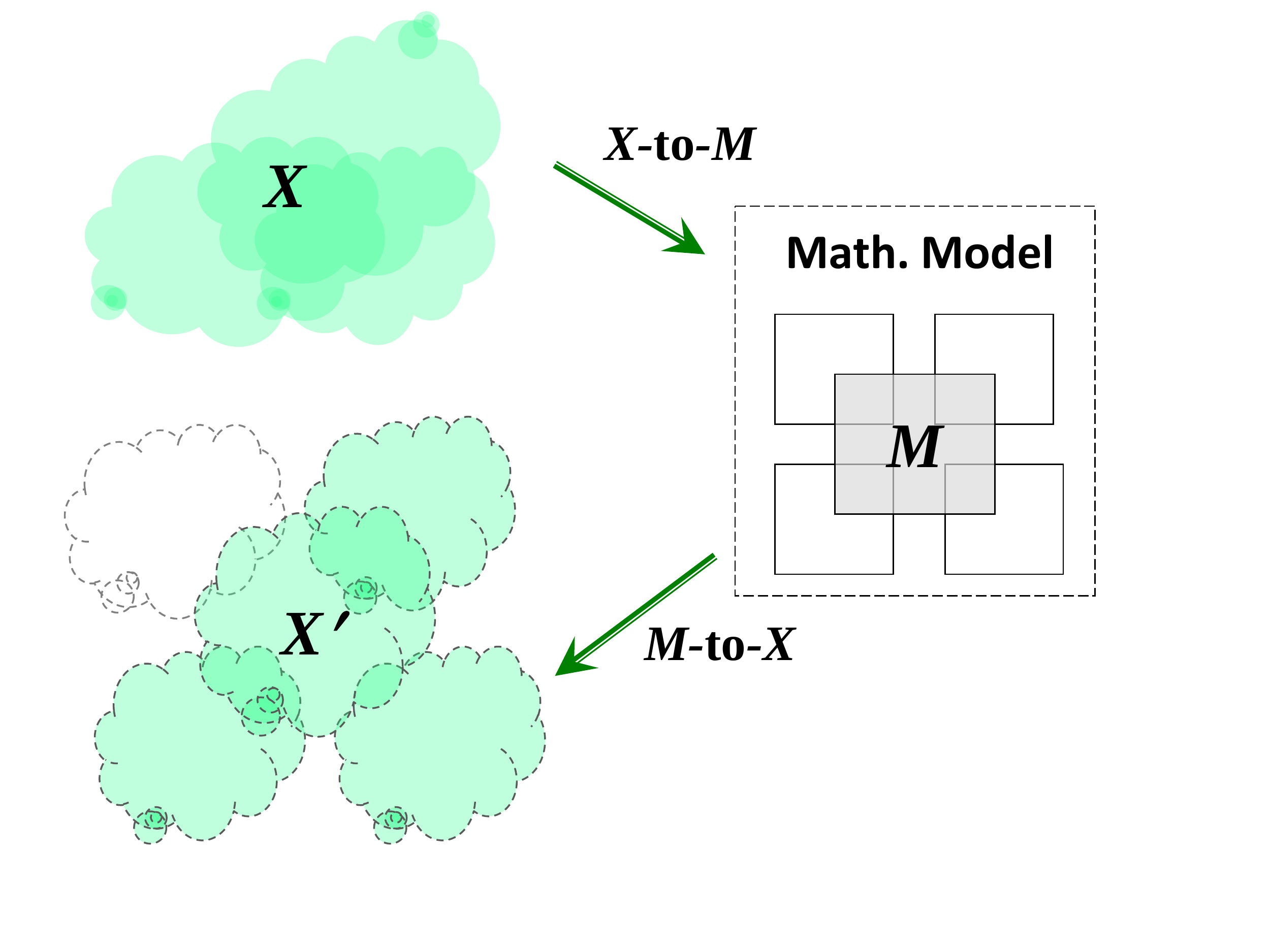}%
\caption{From mathematical models to design patterns}%
\label{fig:soft2math2soft}%
\vspace{-0.2cm}%
\end{wrapfigure}%

Recalling Figure 3, Figure \ref{fig:soft2math2soft} presents a rough
illustration of how mathematical models can reshape our view of a domain
or construct $X$. Building a well-structured mathematical model $M$ of
$X$, and then reinterpreting it back to $X$, can change our view of the
latter as schematically shown in the figure with the reshaped construct
$X'$. Note the discrepancy between the reshaped $X'$ and model $M$:
the upper-left block is missing from $X'$. If $X$ is a piece of reality (think
of mathematical modeling of physical phenomena), this discrepancy
means, \new{most probably,}{} that the model is not adequate \new{(or,
perhaps, some piece of $X$ is not observable).}{} If $X$ is a piece of
software, the discrepancy may point to a deficiency of the design, which
can be fixed by redesigning the software. Even better to base software
design on a well-structured model from the very beginning. Then we say
that model $M$ provides a design pattern for $X$.

We have found several such cases in our work with categorical modeling of
MDE-constructs. For example, the notion of a jointly-monic n-ary arrow
span turns out to be crucial for modeling associations between object
classes, and their correct implementation as well \cite{me-tools08}. It is
interesting to observe how a simple arrow arrangement allows one to
clean the UML metamodel and essentially simplify notation
\cite{me-diagrams02,myTR-assons07}. Another example is modeling
intermodel mappings by Kleisli morphisms, which provide a universal
pattern for model matching (a.k.a alignment) and greatly simplify model
merge as discussed in \sectref{sect-model-merge}. In addition, the Kleisli
view of model mappings provides a design pattern for mapping composition
--- a problem considered to be difficult in the model management
literature \cite{bernstein03}. Sequential composition of symmetric delta
lenses is also not evident; considering such lenses as algebras whose
carriers are profunctors (see Appendix) suggests a precise pattern to be
checked (this work is now in progress). Decomposition of a model
transformation into Cartesian lifting (view execution) followed by the
inverse operation of Cartesian lifting completion (view updating) as
described in Section 4.2.3 provides a useful guidance for model
transformation design, known to be laborious and error-prone. In
particular, it immediately provides bidirectionality.

The graph transformation community also developed several general
patterns applicable to MDE (with models considered as typed attributed
graphs, see \cite{gt-EEPT-2006} for details). In particular, an industrial
standard for model transformation, QVT \cite{QVT08}, was essentially
influenced by triple-graph grammars (TGGs). Some applications of TGGs to
model \syncon\ (and further references) can be found in
\cite{frank-models11}.

\medskip
\mysubsubsection{5.3  Diagrammatic modeling culture and tool
architecture.}

\smallskip
The design patterns mentioned above are based on the respective categorical
machinery (monads, fibrations, profunctors). 
\newok{A software engineer not familiar with these patterns would hardly
recognize them in the arrays of implementation details. Even less probable
is that he will abstract away his implementation concerns and reinvent
such patterns from scratch; distillation of these structures by the CT
community took a good amount of time.}{}  %
In contrast, simple arrow diagrams, like in \figref{fig:bx-scenario}(a) (see
also the Appendix), do not actually need any knowledge of CT: all that is
required is making intermodel relations explicit, and denoting them by
arcs (directed or undirected) connecting the respective objects.  To a
lesser extent, this also holds for the model transformation decomposition
in \figref{fig:trafo-as-getput} and the model merge pattern in
\figref{fig:model-merge}. \new{We refer to a lesser extent because}{}
the former pattern still needs familiarity with the relations-are-spans idea,
and the latter needs an understanding of what colimit is (but, seemingly, it
should be enough to understand it roughly as some
algebraic procedure of ``merging things''). 

The importance of mappings between models/software artifacts is now well
recognized in many communities within SE, and graphical notations have
been employed in SE for a long time. Nevertheless, a majority of model
management tools  neglect the primary status of model mappings: in their
architecture, model matching and alignment are hidden inside
(implementations of) algebraic routines, thus complicating both semantics
and implementation of the latter; concerns are intricately mixed rather
than separated. As all SE textbooks and authorities claim separation of
concerns to be a fundamental principle of software design, an evident
violation of the principle in the cases mentioned above is an empirical fact
that puzzles us.
\del{If model management tools were designed within precise specification
\fwk s based on algebra and logic, we could, perhaps, explain this fact as
follows. A software engineer may well recognize the existence of model
mappings and understand their role in MDE procedures, but as soon as he
needs to present this knowledge in precise algebraic and logical terms, he
at once resorts to familiar discrete patterns of elementwise universal
algebra and ordinary logic. However, MDE-tool design is typically based on
a common SE sense and experience, and it}{} It is not clear why a
BX-tool designer working on tool architecture does not consider simple
arrow diagrams like in \figref{fig:bx-scenario}(a), and prefers discrete
diagrams (b). The latter are, \new{of course,}{} simpler but their
simplicity is deceiving in an almost evident way.

The only explanation we have found is that understanding the deceiving
simplicity of discrete diagrams (b), and, simultaneously,  manageability of
arrow diagrams (a), needs a special diagrammatic modeling culture that a
software engineer normally does not possess. This is the culture of
elementary arrow thinking, which covers the most basic aspects of
manipulating and using arrow diagrams. It appears that even elementary
arrow thinking habits are not cultivated in the current SE curriculum, %
{the corresponding high-level specification patterns are missing from the
software designer toolkit, and software is often structured and modularized
according to the implementation rather than specification concerns.}
%

\section{Related work}

First applications of CT in computer science, and the general claim of CT's
extreme usefulness for computer applications should be, of course,
attributed to Joseph Goguen \cite{GoguenManifest91}. The shift from
modeling semantics of computation (behavior) to modeling structures of
software programs is emphasized by Jos{\'e} Fiadeiro in the introduction
to his book \cite{cats4se}, where he refers to a common ``social'' nature
of both domains. The ideas put forward by Fiadeiro were directly derived
from joint work with Tom Maibaum on what has become known as
component based design and software architecture
\cite{tom92-tempo,tom95-inter,tom95-toolbox}. A clear visualization of
these ideas by \figref{fig:se-x-math} (with M standing for Fiadeiro's
``social'') seems to be new. 
The idea of round-tripping modeling chain \figref{fig:mod-chains} appears
to be novel, its origin can be traced to \cite{me-woodpecker01}.

Don Batory makes an explicit call to using CT in MDE in his invited lecture
for MoDELS'2008 \cite{don-models08}, but he employs the very basic
categorical means, in fact, arrow composition only. In our paper we refer
to much more advanced categorical means: sketches, fibrations, Cartesian
monads, Kleisli categories.

Generalized sketches (graphs with diagram predicates) as a universal
syntactical machinery for formalizing different kinds of models were
proposed by Diskin \etal, \cite{me-diagrams00}. Their application to
special MDE problems can be found in \cite{me-tools08,me-models08-sce}
and in the work of Rutle \etal, see \cite{rutle12},
\cite{rossini12-deepmeta} and references therein. A specific kind of
sketches, ER-sketches, is employed for a number of problems in the
database context by Johnson \etal\ \cite{johnsonERSke}. Considering
models as typed attributed graphs with applications to MDE has been
extensively put forward by the graph transformation (GT) community
\cite{gt-EEPT-2006}; their work is much more operationally oriented than
our concerns in the present paper. On the other hand, in contrast to the
generalized sketches \fwk, constraints seem to be not the first-class
citizens in the GT world.

The shift from functorial to fibrational semantics for sketches to capture
the metamodeling foundations of MDE  was proposed in
\cite{me-dbenchy1-2} and formalized in \cite{me-entcs08}. This
semantics is heavily used in \cite{me-gttse-long},  and in the work of
Rutle \etal\  mentioned above.  Comparison of the two semantic
approaches, functorial and fibrational, and the challenges of proving their
equivalence, are discussed in \cite{uwe07}.


The idea of modeling query languages by monads, and metamodel (or data
schema) mappings by Kleisli mappings, within the functorial semantics
approach, was proposed in \cite{me-adbis97}, and independently by
Johnson and Rosebrugh in their work on ER-sketches
\cite{johnsonERSke}. 
Reformulation of the idea for fibrational semantics was developed and used
for specifying important MDE constructs in
\cite{me-gttse-long,me-mdi10-springer}. An accurate formalization via
Cartesian monads can be found in \cite{me-fase12}. 

Algebraic foundations for BX is now an area of active research. Basics of
the state-based algebraic \fwk\ (lenses) were developed by Pierce with
coauthors \cite{Foster07}; their application to MDE
is due to Stevens \cite{Stevens10-sosym}. Delta-lenses \cite{me-jot11,%
me-models11} is a step towards categorical foundations, but they have
been described in elementary terms using tile algebra
\cite{me-gttse-long}. A categorical approach to the view update problem
has been developed by Johnson and Rosebrugh \etal \cite{Johnson07};
and extended to categorical foundations for lenses based on KZ-monads in
\cite{joro10,joro12}. The notion of symmetric delta lens in Appendix is
new; it results from incorporating the monotonic PutPut-law idea of
Johnson and Rosebrugh into the earlier notion of symmetric delta lens
\cite{me-models11}. Assembling \syncon\ procedures from elementary
blocks is discussed in \cite{me-gttse-long}.

\section{Conclusion}

The paper claims that category theory is a good choice for building
mathematical foundations for MDE. We first discuss two very general
prerequisites that concepts and structures developed in category theory
have to be well applicable for mathematical modeling of MDE-constructs.
We then exemplify the arguments by sketching several categorical models,
which range from general definitions of multimodeling and intermodeling
to important model management scenarios of model merge and
bidirectional update propagation. We \zdmodifyok{show}{briefly explain
(and refer to other work for relevant details)} that these categorical
models provide useful design patterns and guidance for several problems
considered to be difficult.

Moreover, even an elementary arrow arrangement of model merge and BX
scenarios makes explicit a deficiency of the modern tools automating these
scenarios. To wit: these tools' architecture weaves rather than separates
such different concerns as (i) model matching and alignment based on
heuristics and contextual information, and (ii)  relatively simple algebraic
routines of merging and update propagation.  This weaving complicates
both semantics and implementation of the algebraic procedures, does not
allow the user to correct alignment if necessary, and makes tools much
less flexible. It appears that even simple arrow patterns, and the
corresponding structural decisions, may not be evident for a modern
software engineer.

Introduction of CT courses into the SE curriculum, especially in the MDE
context, would be the most natural approach to the problem: even
elementary CT studies should cultivate arrow thinking, develop habits of
diagrammatic reasoning and build a specific intuition of what is a healthy
vs. ill-formed structure. We believe that such intuition, and the structural
lessons one can learn from CT, are of direct relevance for many practical
problems in MDE.

\medskip
\noindent \textbf{Acknowledgement.} Ideas presented in the paper have
been discussed at NECSIS seminars at McMaster and the University of
Waterloo; we are grateful to all their participants, and especially to
Micha{\l} Antkiewicz and  Krzysztof Czarnecki, for useful feedback and
stimulating criticism. We have also benefited greatly from discussions with
Don Batory, Steve Easterbrook, Jos{\'e} Fiadeiro, Michael Healy, Michael
Johnson, Ralf L{\"a}mmel, Bran Selic and Uwe Wolter. Thanks also go to
anonymous referees for comments.

Financial support was provided with the NECSIS project funded by the
Automotive Partnership Canada.

\bibliographystyle{eptcs}
\bibliography{ctse-1,Yrefs}

\begin{thebibliography}{10}
\providecommand{\bibitemdeclare}[2]{}
\providecommand{\surnamestart}{}
\providecommand{\surnameend}{}
\providecommand{\urlprefix}{Available at }
\providecommand{\url}[1]{\texttt{#1}}
\providecommand{\href}[2]{\texttt{#2}}
\providecommand{\urlalt}[2]{\href{#1}{#2}}
\providecommand{\doi}[1]{doi:\urlalt{http://dx.doi.org/#1}{#1}}
\providecommand{\bibinfo}[2]{#2}

\bibitemdeclare{article}{gsdlab-tse09}
\bibitem{gsdlab-tse09}
\bibinfo{author}{Michal \surnamestart Antkiewicz\surnameend},
  \bibinfo{author}{Krzysztof \surnamestart Czarnecki\surnameend} \&
  \bibinfo{author}{Matthew \surnamestart Stephan\surnameend}
  (\bibinfo{year}{2009}): \emph{\bibinfo{title}{Engineering of
  Framework-Specific Modeling Languages}}.
\newblock {\sl \bibinfo{journal}{IEEE Trans. Software Eng.}}
  \bibinfo{volume}{35}(\bibinfo{number}{6}), pp. \bibinfo{pages}{795--824}.
\newblock
  \urlprefix\url{http://doi.ieeecomputersociety.org/10.1109/TSE.2009.30}.

\bibitemdeclare{inproceedings}{don-models08}
\bibitem{don-models08}
\bibinfo{author}{Don~S. \surnamestart Batory\surnameend},
  \bibinfo{author}{Maider \surnamestart Azanza\surnameend} \&
  \bibinfo{author}{Jo{\~a}o \surnamestart Saraiva\surnameend}
  (\bibinfo{year}{2008}): \emph{\bibinfo{title}{The Objects and Arrows of
  Computational Design}}.
\newblock In \bibinfo{editor}{Czarnecki} et~al.  \cite{DBLP:conf/models/2008},
  pp. \bibinfo{pages}{1--20}.
\newblock \urlprefix\url{http://dx.doi.org/10.1007/978-3-540-87875-9_1}.

\bibitemdeclare{inproceedings}{bernstein03}
\bibitem{bernstein03}
\bibinfo{author}{Philip~A. \surnamestart Bernstein\surnameend}
  (\bibinfo{year}{2003}): \emph{\bibinfo{title}{Applying Model Management to
  Classical Meta Data Problems}}.
\newblock In: {\sl \bibinfo{booktitle}{CIDR}}.
\newblock
  \urlprefix\url{http://www-db.cs.wisc.edu/cidr/cidr2003/program/p19.pdf}.

\bibitemdeclare{inproceedings}{Foster08}
\bibitem{Foster08}
\bibinfo{author}{Aaron \surnamestart Bohannon\surnameend},
  \bibinfo{author}{J.~Nathan \surnamestart Foster\surnameend},
  \bibinfo{author}{Benjamin~C. \surnamestart Pierce\surnameend},
  \bibinfo{author}{Alexandre \surnamestart Pilkiewicz\surnameend} \&
  \bibinfo{author}{Alan \surnamestart Schmitt\surnameend}
  (\bibinfo{year}{2008}): \emph{\bibinfo{title}{Boomerang: resourceful lenses
  for string data}}.
\newblock In: {\sl \bibinfo{booktitle}{Proceedings of the 35th annual ACM
  SIGPLAN-SIGACT symposium on Principles of programming languages}},
  \bibinfo{series}{POPL '08}, \bibinfo{publisher}{ACM}, \bibinfo{address}{New
  York, NY, USA}, pp. \bibinfo{pages}{407--419}.
\newblock \urlprefix\url{http://dx.doi.org/10.1145/1328438.1328487}.

\bibitemdeclare{inproceedings}{boronat08}
\bibitem{boronat08}
\bibinfo{author}{Artur \surnamestart Boronat\surnameend},
  \bibinfo{author}{Alexander \surnamestart Knapp\surnameend},
  \bibinfo{author}{Jos{\'e} \surnamestart Meseguer\surnameend} \&
  \bibinfo{author}{Martin \surnamestart Wirsing\surnameend}
  (\bibinfo{year}{2008}): \emph{\bibinfo{title}{What Is a Multi-modeling
  Language?}}
\newblock In \bibinfo{editor}{Andrea \surnamestart Corradini\surnameend} \&
  \bibinfo{editor}{Ugo \surnamestart Montanari\surnameend}, editors: {\sl
  \bibinfo{booktitle}{WADT}}, {\sl \bibinfo{series}{Lecture Notes in Computer
  Science}} \bibinfo{volume}{5486}, \bibinfo{publisher}{Springer}, pp.
  \bibinfo{pages}{71--87}.
\newblock \urlprefix\url{http://dx.doi.org/10.1007/978-3-642-03429-9_6}.

\bibitemdeclare{inproceedings}{Czarnecki09a}
\bibitem{Czarnecki09a}
\bibinfo{author}{Krzysztof \surnamestart Czarnecki\surnameend},
  \bibinfo{author}{J.~Nathan \surnamestart Foster\surnameend},
  \bibinfo{author}{Zhenjiang \surnamestart Hu\surnameend},
  \bibinfo{author}{Ralf \surnamestart L{\"a}mmel\surnameend},
  \bibinfo{author}{Andy \surnamestart Sch{\"u}rr\surnameend} \&
  \bibinfo{author}{James~F. \surnamestart Terwilliger\surnameend}
  (\bibinfo{year}{2009}): \emph{\bibinfo{title}{Bidirectional Transformations:
  A Cross-Discipline Perspective}}.
\newblock In \bibinfo{editor}{Richard~F. \surnamestart Paige\surnameend},
  editor: {\sl \bibinfo{booktitle}{ICMT}}, {\sl \bibinfo{series}{Lecture Notes
  in Computer Science}} \bibinfo{volume}{5563}, \bibinfo{publisher}{Springer},
  pp. \bibinfo{pages}{260--283}.
\newblock \urlprefix\url{http://dx.doi.org/10.1007/978-3-642-02408-5_19}.

\bibitemdeclare{proceedings}{DBLP:conf/models/2008}
\bibitem{DBLP:conf/models/2008}
\bibinfo{editor}{Krzysztof \surnamestart Czarnecki\surnameend},
  \bibinfo{editor}{Ileana \surnamestart Ober\surnameend},
  \bibinfo{editor}{Jean-Michel \surnamestart Bruel\surnameend},
  \bibinfo{editor}{Axel \surnamestart Uhl\surnameend} \&
  \bibinfo{editor}{Markus \surnamestart V{\"o}lter\surnameend}, editors
  (\bibinfo{year}{2008}): \emph{\bibinfo{title}{Model Driven Engineering
  Languages and Systems, 11th International Conference, MoDELS 2008, Toulouse,
  France, September 28 - October 3, 2008. Proceedings}}. {\sl
  \bibinfo{series}{Lecture Notes in Computer Science}} \bibinfo{volume}{5301},
  \bibinfo{publisher}{Springer}.
\newblock \urlprefix\url{http://dx.doi.org/10.1007/978-3-540-87875-9}.

\bibitemdeclare{techreport}{myTR-assons07}
\bibitem{myTR-assons07}
\bibinfo{author}{Z.~\surnamestart Diskin\surnameend} (\bibinfo{year}{2007}):
  \emph{\bibinfo{title}{{Mappings, maps, atlases and tables: A formal semantics
  for associations in UML2}}}.
\newblock \bibinfo{type}{Technical Report} \bibinfo{number}{CSRG-566},
  \bibinfo{institution}{{University of Toronto}}.
\newblock
  \bibinfo{note}{\url{http://ftp.cs.toronto.edu/pub/reports/csrg/566/TR-566-umlAssons.pdf}}.

\bibitemdeclare{techreport}{myTR-integr08}
\bibitem{myTR-integr08}
\bibinfo{author}{Z.~\surnamestart Diskin\surnameend},
  \bibinfo{author}{S.~\surnamestart Easterbrook\surnameend} \&
  \bibinfo{author}{R.~\surnamestart Miller\surnameend} (\bibinfo{year}{2008}):
  \emph{\bibinfo{title}{Integrating schema integration frameworks,
  algebraically}}.
\newblock \bibinfo{type}{Technical Report} \bibinfo{number}{{CSRG-583}},
  \bibinfo{institution}{{University of Toronto}}.
\newblock
  \bibinfo{note}{\url{http://ftp.cs.toronto.edu/pub/reports/csrg/583/TR-583-schemaIntegr.pdf}}.

\bibitemdeclare{inproceedings}{me-models11}
\bibitem{me-models11}
\bibinfo{author}{Z.~\surnamestart Diskin\surnameend},
  \bibinfo{author}{Y.~\surnamestart Xiong\surnameend},
  \bibinfo{author}{K.~\surnamestart Czarnecki\surnameend},
  \bibinfo{author}{H.~\surnamestart Ehrig\surnameend},
  \bibinfo{author}{F.~\surnamestart Hermann\surnameend} \&
  \bibinfo{author}{F.~\surnamestart Orejas\surnameend} (\bibinfo{year}{2011}):
  \emph{\bibinfo{title}{From State- to Delta-Based Bidirectional Model
  Transformations: The Symmetric Case}}.
\newblock In \bibinfo{editor}{Whittle} et~al.  \cite{DBLP:conf/models/2011},
  pp. \bibinfo{pages}{304--318}.
\newblock \urlprefix\url{http://dx.doi.org/10.1007/978-3-642-24485-8_22}.

\bibitemdeclare{inproceedings}{me-woodpecker01}
\bibitem{me-woodpecker01}
\bibinfo{author}{Zinovy \surnamestart Diskin\surnameend}
  (\bibinfo{year}{2001}): \emph{\bibinfo{title}{On Modeling, Mathematics,
  Category Theory and RM-ODP}}.
\newblock In \bibinfo{editor}{Jos{\'e} A.~Moinhos \surnamestart
  Cordeiro\surnameend} \& \bibinfo{editor}{Haim \surnamestart
  Kilov\surnameend}, editors: {\sl \bibinfo{booktitle}{WOODPECKER}},
  \bibinfo{publisher}{ICEIS Press}, pp. \bibinfo{pages}{38--54}.

\bibitemdeclare{inproceedings}{me-diagrams02}
\bibitem{me-diagrams02}
\bibinfo{author}{Zinovy \surnamestart Diskin\surnameend}
  (\bibinfo{year}{2002}): \emph{\bibinfo{title}{Visualization vs. Specification
  in Diagrammatic Notations: A Case Study with the UML}}.
\newblock In \bibinfo{editor}{Mary \surnamestart Hegarty\surnameend},
  \bibinfo{editor}{Bernd \surnamestart Meyer\surnameend} \&
  \bibinfo{editor}{N.~Hari \surnamestart Narayanan\surnameend}, editors: {\sl
  \bibinfo{booktitle}{Diagrams}}, {\sl \bibinfo{series}{Lecture Notes in
  Computer Science}} \bibinfo{volume}{2317}, \bibinfo{publisher}{Springer}, pp.
  \bibinfo{pages}{112--115}.
\newblock \urlprefix\url{http://dx.doi.org/10.1007/3-540-46037-3_15}.

\bibitemdeclare{incollection}{me-dbenchy1-2}
\bibitem{me-dbenchy1-2}
\bibinfo{author}{Zinovy \surnamestart Diskin\surnameend}
  (\bibinfo{year}{2005}): \emph{\bibinfo{title}{Mathematics of Generic
  Specifications for Model Management}}.
\newblock In \bibinfo{editor}{Laura~C. \surnamestart Rivero\surnameend},
  \bibinfo{editor}{Jorge~Horacio \surnamestart Doorn\surnameend} \&
  \bibinfo{editor}{Viviana~E. \surnamestart Ferraggine\surnameend}, editors:
  {\sl \bibinfo{booktitle}{Encyclopedia of Database Technologies and
  Applications}}, \bibinfo{publisher}{Idea Group}, pp.
  \bibinfo{pages}{351--366}.

\bibitemdeclare{inproceedings}{me-models08}
\bibitem{me-models08}
\bibinfo{author}{Zinovy \surnamestart Diskin\surnameend}
  (\bibinfo{year}{2008}): \emph{\bibinfo{title}{Algebraic Models for
  Bidirectional Model Synchronization}}.
\newblock In \bibinfo{editor}{Czarnecki} et~al.  \cite{DBLP:conf/models/2008},
  pp. \bibinfo{pages}{21--36}.
\newblock \urlprefix\url{http://dx.doi.org/10.1007/978-3-540-87875-9_2}.

\bibitemdeclare{inproceedings}{me-gttse-long}
\bibitem{me-gttse-long}
\bibinfo{author}{Zinovy \surnamestart Diskin\surnameend}
  (\bibinfo{year}{2009}): \emph{\bibinfo{title}{Model Synchronization:
  Mappings, Tiles, and Categories}}.
\newblock In \bibinfo{editor}{Jo{\~a}o~M. \surnamestart Fernandes\surnameend},
  \bibinfo{editor}{Ralf \surnamestart L{\"a}mmel\surnameend},
  \bibinfo{editor}{Joost \surnamestart Visser\surnameend} \&
  \bibinfo{editor}{Jo{\~a}o \surnamestart Saraiva\surnameend}, editors: {\sl
  \bibinfo{booktitle}{GTTSE}}, {\sl \bibinfo{series}{Lecture Notes in Computer
  Science}} \bibinfo{volume}{6491}, \bibinfo{publisher}{Springer}, pp.
  \bibinfo{pages}{92--165}.
\newblock \urlprefix\url{http://dx.doi.org/10.1007/978-3-642-18023-1_3}.

\bibitemdeclare{inproceedings}{me-adbis97}
\bibitem{me-adbis97}
\bibinfo{author}{Zinovy \surnamestart Diskin\surnameend} \&
  \bibinfo{author}{Boris \surnamestart Cadish\surnameend}
  (\bibinfo{year}{1997}): \emph{\bibinfo{title}{A Graphical Yet Formalized
  Framework for Specifying View Systems}}.
\newblock In: {\sl \bibinfo{booktitle}{ADBIS}}, \bibinfo{publisher}{Nevsky
  Dialect}, pp. \bibinfo{pages}{123--132}.
\newblock \urlprefix\url{http://www.bcs.org/upload/pdf/ewic_ad97_paper17.pdf}.

\bibitemdeclare{inproceedings}{me-tools08}
\bibitem{me-tools08}
\bibinfo{author}{Zinovy \surnamestart Diskin\surnameend},
  \bibinfo{author}{Steve~M. \surnamestart Easterbrook\surnameend} \&
  \bibinfo{author}{J{\"u}rgen \surnamestart Dingel\surnameend}
  (\bibinfo{year}{2008}): \emph{\bibinfo{title}{Engineering Associations: From
  Models to Code and Back through Semantics}}.
\newblock In \bibinfo{editor}{Richard~F. \surnamestart Paige\surnameend} \&
  \bibinfo{editor}{Bertrand \surnamestart Meyer\surnameend}, editors: {\sl
  \bibinfo{booktitle}{TOOLS (46)}}, {\sl \bibinfo{series}{Lecture Notes in
  Business Information Processing}}~\bibinfo{volume}{11},
  \bibinfo{publisher}{Springer}, pp. \bibinfo{pages}{336--355}.
\newblock \urlprefix\url{http://dx.doi.org/10.1007/978-3-540-69824-1_19}.

\bibitemdeclare{inproceedings}{me-diagrams00}
\bibitem{me-diagrams00}
\bibinfo{author}{Zinovy \surnamestart Diskin\surnameend},
  \bibinfo{author}{Boris \surnamestart Kadish\surnameend},
  \bibinfo{author}{Frank \surnamestart Piessens\surnameend} \&
  \bibinfo{author}{Michael \surnamestart Johnson\surnameend}
  (\bibinfo{year}{2000}): \emph{\bibinfo{title}{Universal Arrow Foundations for
  Visual Modeling}}.
\newblock In \bibinfo{editor}{Michael \surnamestart Anderson\surnameend},
  \bibinfo{editor}{Peter \surnamestart Cheng\surnameend} \&
  \bibinfo{editor}{Volker \surnamestart Haarslev\surnameend}, editors: {\sl
  \bibinfo{booktitle}{Diagrams}}, {\sl \bibinfo{series}{Lecture Notes in
  Computer Science}} \bibinfo{volume}{1889}, \bibinfo{publisher}{Springer}, pp.
  \bibinfo{pages}{345--360}.
\newblock
  \urlprefix\url{http://link.springer.de/link/service/series/0558/bibs/1889/18890345.htm}.

\bibitemdeclare{inproceedings}{me-fase12}
\bibitem{me-fase12}
\bibinfo{author}{Zinovy \surnamestart Diskin\surnameend}, \bibinfo{author}{Tom
  \surnamestart Maibaum\surnameend} \& \bibinfo{author}{Krzysztof \surnamestart
  Czarnecki\surnameend} (\bibinfo{year}{2012}):
  \emph{\bibinfo{title}{Intermodeling, Queries, and Kleisli Categories}}.
\newblock In \bibinfo{editor}{Juan \surnamestart de~Lara\surnameend} \&
  \bibinfo{editor}{Andrea \surnamestart Zisman\surnameend}, editors: {\sl
  \bibinfo{booktitle}{FASE}}, {\sl \bibinfo{series}{Lecture Notes in Computer
  Science}} \bibinfo{volume}{7212}, \bibinfo{publisher}{Springer}, pp.
  \bibinfo{pages}{163--177}.
\newblock \urlprefix\url{http://dx.doi.org/10.1007/978-3-642-28872-2_12}.

\bibitemdeclare{article}{me-entcs08}
\bibitem{me-entcs08}
\bibinfo{author}{Zinovy \surnamestart Diskin\surnameend} \&
  \bibinfo{author}{Uwe \surnamestart Wolter\surnameend} (\bibinfo{year}{2008}):
  \emph{\bibinfo{title}{A Diagrammatic Logic for Object-Oriented Visual
  Modeling}}.
\newblock {\sl \bibinfo{journal}{Electr. Notes Theor. Comput. Sci.}}
  \bibinfo{volume}{203}(\bibinfo{number}{6}), pp. \bibinfo{pages}{19--41}.
\newblock \urlprefix\url{http://dx.doi.org/10.1016/j.entcs.2008.10.041}.

\bibitemdeclare{inproceedings}{me-mdi10-springer}
\bibitem{me-mdi10-springer}
\bibinfo{author}{Zinovy \surnamestart Diskin\surnameend},
  \bibinfo{author}{Yingfei \surnamestart Xiong\surnameend} \&
  \bibinfo{author}{Krzysztof \surnamestart Czarnecki\surnameend}
  (\bibinfo{year}{2010}): \emph{\bibinfo{title}{Specifying Overlaps of
  Heterogeneous Models for Global Consistency Checking}}.
\newblock In: {\sl \bibinfo{booktitle}{MoDELS Workshops}}, {\sl
  \bibinfo{series}{Lecture Notes in Computer Science}} \bibinfo{volume}{6627},
  \bibinfo{publisher}{Springer}, pp. \bibinfo{pages}{165--179}.
\newblock \urlprefix\url{http://dx.doi.org/10.1007/978-3-642-21210-9_16}.

\bibitemdeclare{article}{me-jot11}
\bibitem{me-jot11}
\bibinfo{author}{Zinovy \surnamestart Diskin\surnameend},
  \bibinfo{author}{Yingfei \surnamestart Xiong\surnameend} \&
  \bibinfo{author}{Krzysztof \surnamestart Czarnecki\surnameend}
  (\bibinfo{year}{2011}): \emph{\bibinfo{title}{From State- to Delta-Based
  Bidirectional Model Transformations: the Asymmetric Case}}.
\newblock {\sl \bibinfo{journal}{Journal of Object Technology}}
  \bibinfo{volume}{10}, pp. \bibinfo{pages}{6: 1--25}.
\newblock \urlprefix\url{http://dx.doi.org/10.5381/jot.2011.10.1.a6}.

\bibitemdeclare{book}{gt-EEPT-2006}
\bibitem{gt-EEPT-2006}
\bibinfo{author}{H.~\surnamestart Ehrig\surnameend},
  \bibinfo{author}{K.~\surnamestart Ehrig\surnameend},
  \bibinfo{author}{U.~\surnamestart Prange\surnameend} \&
  \bibinfo{author}{G.~\surnamestart Taenzer\surnameend} (\bibinfo{year}{2006}):
  \emph{\bibinfo{title}{{Fundamentals of Algebraic Graph Transformation}}}.

\bibitemdeclare{inproceedings}{tom95-toolbox}
\bibitem{tom95-toolbox}
\bibinfo{author}{J.~L. \surnamestart Fiadeiro\surnameend} \&
  \bibinfo{author}{T.~S.~E. \surnamestart Maibaum\surnameend}
  (\bibinfo{year}{1995}): \emph{\bibinfo{title}{A Mathematical Toolbox for the
  Software Architect}}.
\newblock In \bibinfo{editor}{J.~\surnamestart Kramer\surnameend} \&
  \bibinfo{editor}{A.~\surnamestart Wolf\surnameend}, editors: {\sl
  \bibinfo{booktitle}{8th Int. Workshop on Software Specification and Design}},
  \bibinfo{publisher}{IEEE CS Press}, pp. \bibinfo{pages}{46--55}.

\bibitemdeclare{inproceedings}{fiadeiro04}
\bibitem{fiadeiro04}
\bibinfo{author}{Jos{\'e}~Luiz \surnamestart Fiadeiro\surnameend}
  (\bibinfo{year}{2004}): \emph{\bibinfo{title}{Software Services: Scientific
  Challenge or Industrial Hype?}}
\newblock In \bibinfo{editor}{Zhiming \surnamestart Liu\surnameend} \&
  \bibinfo{editor}{Keijiro \surnamestart Araki\surnameend}, editors: {\sl
  \bibinfo{booktitle}{ICTAC}}, {\sl \bibinfo{series}{Lecture Notes in Computer
  Science}} \bibinfo{volume}{3407}, \bibinfo{publisher}{Springer}, pp.
  \bibinfo{pages}{1--13}.
\newblock \urlprefix\url{http://dx.doi.org/10.1007/978-3-540-31862-0_1}.

\bibitemdeclare{article}{tom92-tempo}
\bibitem{tom92-tempo}
\bibinfo{author}{Jos{\'e}~Luiz \surnamestart Fiadeiro\surnameend} \&
  \bibinfo{author}{T.~S.~E. \surnamestart Maibaum\surnameend}
  (\bibinfo{year}{1992}): \emph{\bibinfo{title}{Temporal Theories as
  Modularisation Units for Concurrent System Specification}}.
\newblock {\sl \bibinfo{journal}{Formal Asp. Comput.}}
  \bibinfo{volume}{4}(\bibinfo{number}{3}), pp. \bibinfo{pages}{239--272}.
\newblock \urlprefix\url{http://dx.doi.org/10.1007/BF01212304}.

\bibitemdeclare{inproceedings}{tom95-inter}
\bibitem{tom95-inter}
\bibinfo{author}{Jos{\'e}~Luiz \surnamestart Fiadeiro\surnameend} \&
  \bibinfo{author}{T.~S.~E. \surnamestart Maibaum\surnameend}
  (\bibinfo{year}{1995}): \emph{\bibinfo{title}{Interconnecting Formalisms:
  Supporting Modularity, Reuse and Incrementality}}.
\newblock In: {\sl \bibinfo{booktitle}{SIGSOFT FSE}}, pp.
  \bibinfo{pages}{72--80}.
\newblock \urlprefix\url{http://doi.acm.org/10.1145/222124.222141}.

\bibitemdeclare{article}{Foster07}
\bibitem{Foster07}
\bibinfo{author}{J.~N. \surnamestart Foster\surnameend},
  \bibinfo{author}{M.~\surnamestart Greenwald\surnameend},
  \bibinfo{author}{J.~\surnamestart Moore\surnameend},
  \bibinfo{author}{B.~\surnamestart Pierce\surnameend} \&
  \bibinfo{author}{A.~\surnamestart Schmitt\surnameend} (\bibinfo{year}{2007}):
  \emph{\bibinfo{title}{Combinators for bidirectional tree transformations: A
  linguistic approach to the view-update problem}}.
\newblock {\sl \bibinfo{journal}{ACM Trans. Program. Lang. Syst.}}
  \bibinfo{volume}{29}(\bibinfo{number}{3}), \doi{10.1145/1232420.1232424}.

\bibitemdeclare{article}{GoguenManifest91}
\bibitem{GoguenManifest91}
\bibinfo{author}{Joseph~A. \surnamestart Goguen\surnameend}
  (\bibinfo{year}{1991}): \emph{\bibinfo{title}{A Categorical Manifesto}}.
\newblock {\sl \bibinfo{journal}{Mathematical Structures in Computer Science}}
  \bibinfo{volume}{1}(\bibinfo{number}{1}), pp. \bibinfo{pages}{49--67}.
\newblock \urlprefix\url{http://dx.doi.org/10.1017/S0960129500000050}.

\bibitemdeclare{inproceedings}{frank-models11}
\bibitem{frank-models11}
\bibinfo{author}{Frank \surnamestart Hermann\surnameend},
  \bibinfo{author}{Hartmut \surnamestart Ehrig\surnameend},
  \bibinfo{author}{Fernando \surnamestart Orejas\surnameend},
  \bibinfo{author}{Krzysztof \surnamestart Czarnecki\surnameend},
  \bibinfo{author}{Zinovy \surnamestart Diskin\surnameend} \&
  \bibinfo{author}{Yingfei \surnamestart Xiong\surnameend}
  (\bibinfo{year}{2011}): \emph{\bibinfo{title}{Correctness of Model
  Synchronization Based on Triple Graph Grammars}}.
\newblock In \bibinfo{editor}{Whittle} et~al.  \cite{DBLP:conf/models/2011},
  pp. \bibinfo{pages}{668--682}.
\newblock \urlprefix\url{http://dx.doi.org/10.1007/978-3-642-24485-8_49}.

\bibitemdeclare{inproceedings}{Hofmann11}
\bibitem{Hofmann11}
\bibinfo{author}{M.~\surnamestart Hofmann\surnameend},
  \bibinfo{author}{B.~\surnamestart Pierce\surnameend} \&
  \bibinfo{author}{D.~\surnamestart Wagner\surnameend} (\bibinfo{year}{2011}):
  \emph{\bibinfo{title}{Symmetric Lenses}}.
\newblock In: {\sl \bibinfo{booktitle}{POPL}}.
\newblock \urlprefix\url{http://doi.acm.org/10.1145/1328438.1328487}.

\bibitemdeclare{article}{johnsonERSke}
\bibitem{johnsonERSke}
\bibinfo{author}{M.~\surnamestart Johnson\surnameend},
  \bibinfo{author}{R.~\surnamestart Rosebrugh\surnameend} \&
  \bibinfo{author}{R.~\surnamestart Wood\surnameend} (\bibinfo{year}{2002}):
  \emph{\bibinfo{title}{Entity-relationship-attribute designs and sketches}}.
\newblock {\sl \bibinfo{journal}{Theory and Applications of Categories}}
  \bibinfo{volume}{10}(\bibinfo{number}{3}), pp. \bibinfo{pages}{94--112}.

\bibitemdeclare{article}{Johnson07}
\bibitem{Johnson07}
\bibinfo{author}{Michael \surnamestart Johnson\surnameend} \&
  \bibinfo{author}{Robert~D. \surnamestart Rosebrugh\surnameend}
  (\bibinfo{year}{2007}): \emph{\bibinfo{title}{Fibrations and universal view
  updatability}}.
\newblock {\sl \bibinfo{journal}{Theor. Comput. Sci.}}
  \bibinfo{volume}{388}(\bibinfo{number}{1-3}), pp. \bibinfo{pages}{109--129}.
\newblock \urlprefix\url{http://dx.doi.org/10.1016/j.tcs.2007.06.004}.

\bibitemdeclare{inproceedings}{joro12}
\bibitem{joro12}
\bibinfo{author}{Michael \surnamestart Johnson\surnameend} \&
  \bibinfo{author}{Robert~D. \surnamestart Rosebrugh\surnameend}
  (\bibinfo{year}{2012}): \emph{\bibinfo{title}{Lens put-put laws: monotonic
  and mixed}}.
\newblock \bibinfo{note}{To appear. http://www.easst.org/eceasst}.

\bibitemdeclare{article}{joro10}
\bibitem{joro10}
\bibinfo{author}{Michael \surnamestart Johnson\surnameend},
  \bibinfo{author}{Robert~D. \surnamestart Rosebrugh\surnameend} \&
  \bibinfo{author}{Richard~J. \surnamestart Wood\surnameend}
  (\bibinfo{year}{2010}): \emph{\bibinfo{title}{Algebras and Update
  Strategies}}.
\newblock {\sl \bibinfo{journal}{J. UCS}}
  \bibinfo{volume}{16}(\bibinfo{number}{5}), pp. \bibinfo{pages}{729--748}.
\newblock \urlprefix\url{http://dx.doi.org/10.3217/jucs-016-05-0729}.

\bibitemdeclare{book}{cats4se}
\bibitem{cats4se}
\bibinfo{author}{\surnamestart {Jos\'e Fiadeiro}\surnameend}
  (\bibinfo{year}{2004}): \emph{\bibinfo{title}{Categories for Software
  Engineering}}.
\newblock \bibinfo{publisher}{Springer}.

\bibitemdeclare{inproceedings}{gabi-models09}
\bibitem{gabi-models09}
\bibinfo{author}{Stefan \surnamestart Jurack\surnameend} \&
  \bibinfo{author}{Gabriele \surnamestart Taentzer\surnameend}
  (\bibinfo{year}{2009}): \emph{\bibinfo{title}{Towards Composite Model
  Transformations Using Distributed Graph Transformation Concepts}}.
\newblock In \bibinfo{editor}{Andy \surnamestart Sch{\"u}rr\surnameend} \&
  \bibinfo{editor}{Bran \surnamestart Selic\surnameend}, editors: {\sl
  \bibinfo{booktitle}{MoDELS}}, {\sl \bibinfo{series}{Lecture Notes in Computer
  Science}} \bibinfo{volume}{5795}, \bibinfo{publisher}{Springer}, pp.
  \bibinfo{pages}{226--240}.
\newblock \urlprefix\url{http://dx.doi.org/10.1007/978-3-642-04425-0_17}.

\bibitemdeclare{inproceedings}{me-models08-sce}
\bibitem{me-models08-sce}
\bibinfo{author}{Hongzhi \surnamestart Liang\surnameend},
  \bibinfo{author}{Zinovy \surnamestart Diskin\surnameend},
  \bibinfo{author}{J{\"u}rgen \surnamestart Dingel\surnameend} \&
  \bibinfo{author}{Ernesto \surnamestart Posse\surnameend}
  (\bibinfo{year}{2008}): \emph{\bibinfo{title}{A General Approach for Scenario
  Integration}}.
\newblock In \bibinfo{editor}{Czarnecki} et~al.  \cite{DBLP:conf/models/2008},
  pp. \bibinfo{pages}{204--218}.
\newblock \urlprefix\url{http://dx.doi.org/10.1007/978-3-540-87875-9_15}.

\bibitemdeclare{article}{Makkai97}
\bibitem{Makkai97}
\bibinfo{author}{M.~\surnamestart Makkai\surnameend} (\bibinfo{year}{1997}):
  \emph{\bibinfo{title}{Generalized Sketches as a Framework for Completeness
  Theorems}}.
\newblock {\sl \bibinfo{journal}{Journal of Pure and Applied Algebra}}
  \bibinfo{volume}{115}, pp. \bibinfo{pages}{49--79, 179--212, 214--274}.

\bibitemdeclare{inproceedings}{Matsuda07}
\bibitem{Matsuda07}
\bibinfo{author}{Kazutaka \surnamestart Matsuda\surnameend},
  \bibinfo{author}{Zhenjiang \surnamestart Hu\surnameend},
  \bibinfo{author}{Keisuke \surnamestart Nakano\surnameend},
  \bibinfo{author}{Makoto \surnamestart Hamana\surnameend} \&
  \bibinfo{author}{Masato \surnamestart Takeichi\surnameend}
  (\bibinfo{year}{2007}): \emph{\bibinfo{title}{Bidirectionalization
  transformation based on automatic derivation of view complement functions}}.
\newblock In \bibinfo{editor}{Ralf \surnamestart Hinze\surnameend} \&
  \bibinfo{editor}{Norman \surnamestart Ramsey\surnameend}, editors: {\sl
  \bibinfo{booktitle}{ICFP}}, \bibinfo{publisher}{ACM}, pp.
  \bibinfo{pages}{47--58}.
\newblock \urlprefix\url{http://dx.doi.org/10.1145/1291151.1291162}.

\bibitemdeclare{misc}{QVT08}
\bibitem{QVT08}
\bibinfo{author}{\surnamestart {Object Management Group}\surnameend}
  (\bibinfo{year}{2008}): \emph{\bibinfo{title}{{MOF} Query / Views /
  Transformations Specification 1.0}}.
\newblock
  \bibinfo{howpublished}{\url{http://www.omg.org/docs/formal/08-04-03.pdf}}.

\bibitemdeclare{inproceedings}{pottinger03}
\bibitem{pottinger03}
\bibinfo{author}{Rachel \surnamestart Pottinger\surnameend} \&
  \bibinfo{author}{Philip~A. \surnamestart Bernstein\surnameend}
  (\bibinfo{year}{2003}): \emph{\bibinfo{title}{Merging Models Based on Given
  Correspondences}}.
\newblock In: {\sl \bibinfo{booktitle}{VLDB}}, pp. \bibinfo{pages}{826--873}.
\newblock \urlprefix\url{http://www.vldb.org/conf/2003/papers/S26P01.pdf}.

\bibitemdeclare{inproceedings}{rossini12-deepmeta}
\bibitem{rossini12-deepmeta}
\bibinfo{author}{Alessandro \surnamestart Rossini\surnameend},
  \bibinfo{author}{\surnamestart \surnameend}, \bibinfo{author}{Juan
  \surnamestart de~Lara\surnameend}, \bibinfo{author}{Esther \surnamestart
  Guerra\surnameend}, \bibinfo{author}{Adrian \surnamestart Rutle\surnameend}
  \& \bibinfo{author}{Yngve \surnamestart Lamo\surnameend}
  (\bibinfo{year}{2012}): \emph{\bibinfo{title}{A Graph Transformation-based
  Semantics for Deep Metamodelling}}.
\newblock In: {\sl \bibinfo{booktitle}{AGTIVE 2012}}.
\newblock \urlprefix\url{http://dx.doi.org/10.1016/j.jlap.2009.10.003}.
\newblock \bibinfo{note}{To appear}.

\bibitemdeclare{article}{rossini10}
\bibitem{rossini10}
\bibinfo{author}{Alessandro \surnamestart Rossini\surnameend},
  \bibinfo{author}{Adrian \surnamestart Rutle\surnameend},
  \bibinfo{author}{Yngve \surnamestart Lamo\surnameend} \& \bibinfo{author}{Uwe
  \surnamestart Wolter\surnameend} (\bibinfo{year}{2010}):
  \emph{\bibinfo{title}{A formalisation of the copy-modify-merge approach to
  version control in MDE}}.
\newblock {\sl \bibinfo{journal}{J. Log. Algebr. Program.}}
  \bibinfo{volume}{79}(\bibinfo{number}{7}), pp. \bibinfo{pages}{636--658}.
\newblock \urlprefix\url{http://dx.doi.org/10.1016/j.jlap.2009.10.003}.

\bibitemdeclare{inproceedings}{rutle-fase10}
\bibitem{rutle-fase10}
\bibinfo{author}{Adrian \surnamestart Rutle\surnameend},
  \bibinfo{author}{Alessandro \surnamestart Rossini\surnameend},
  \bibinfo{author}{Yngve \surnamestart Lamo\surnameend} \& \bibinfo{author}{Uwe
  \surnamestart Wolter\surnameend} (\bibinfo{year}{2010}):
  \emph{\bibinfo{title}{A Formalisation of Constraint-Aware Model
  Transformations}}.
\newblock In \bibinfo{editor}{David~S. \surnamestart Rosenblum\surnameend} \&
  \bibinfo{editor}{Gabriele \surnamestart Taentzer\surnameend}, editors: {\sl
  \bibinfo{booktitle}{FASE}}, {\sl \bibinfo{series}{Lecture Notes in Computer
  Science}} \bibinfo{volume}{6013}, \bibinfo{publisher}{Springer}, pp.
  \bibinfo{pages}{13--28}.
\newblock \urlprefix\url{http://dx.doi.org/10.1007/978-3-642-12029-9_2}.

\bibitemdeclare{article}{rutle12}
\bibitem{rutle12}
\bibinfo{author}{Adrian \surnamestart Rutle\surnameend},
  \bibinfo{author}{Alessandro \surnamestart Rossini\surnameend},
  \bibinfo{author}{Yngve \surnamestart Lamo\surnameend} \& \bibinfo{author}{Uwe
  \surnamestart Wolter\surnameend} (\bibinfo{year}{2012}):
  \emph{\bibinfo{title}{A formal approach to the specification and
  transformation of constraints in MDE}}.
\newblock {\sl \bibinfo{journal}{J. Log. Algebr. Program.}}
  \bibinfo{volume}{81}(\bibinfo{number}{4}), pp. \bibinfo{pages}{422--457}.
\newblock \urlprefix\url{http://dx.doi.org/10.1016/j.jlap.2012.03.006}.

\bibitemdeclare{article}{bran08}
\bibitem{bran08}
\bibinfo{author}{Bran \surnamestart Selic\surnameend} (\bibinfo{year}{2008}):
  \emph{\bibinfo{title}{Personal reflections on automation, programming
  culture, and model-based software engineering}}.
\newblock {\sl \bibinfo{journal}{Autom. Softw. Eng.}}
  \bibinfo{volume}{15}(\bibinfo{number}{3-4}), pp. \bibinfo{pages}{379--391}.
\newblock \urlprefix\url{http://dx.doi.org/10.1007/s10515-008-0035-7}.

\bibitemdeclare{inproceedings}{shaw96}
\bibitem{shaw96}
\bibinfo{author}{M.~\surnamestart Shaw\surnameend} (\bibinfo{year}{1996}):
  \emph{\bibinfo{title}{Three patterns that help explain the development of
  software engineering (position paper)}}.
\newblock In: {\sl \bibinfo{booktitle}{{Dagstuhl Workshop on Software
  Architecture}}}.

\bibitemdeclare{article}{spacca94}
\bibitem{spacca94}
\bibinfo{author}{Stefano \surnamestart Spaccapietra\surnameend} \&
  \bibinfo{author}{Christine \surnamestart Parent\surnameend}
  (\bibinfo{year}{1994}): \emph{\bibinfo{title}{View Integration: A Step
  Forward in Solving Structural Conflicts}}.
\newblock {\sl \bibinfo{journal}{IEEE Trans. Knowl. Data Eng.}}
  \bibinfo{volume}{6}(\bibinfo{number}{2}), pp. \bibinfo{pages}{258--274}.
\newblock \urlprefix\url{http://doi.ieeecomputersociety.org/10.1109/69.277770}.

\bibitemdeclare{article}{Stevens10-sosym}
\bibitem{Stevens10-sosym}
\bibinfo{author}{Perdita \surnamestart Stevens\surnameend}
  (\bibinfo{year}{2010}): \emph{\bibinfo{title}{Bidirectional model
  transformations in QVT: semantic issues and open questions}}.
\newblock {\sl \bibinfo{journal}{Software and System Modeling}}
  \bibinfo{volume}{9}(\bibinfo{number}{1}), pp. \bibinfo{pages}{7--20}.
\newblock \urlprefix\url{http://dx.doi.org/10.1007/s10270-008-0109-9}.

\bibitemdeclare{proceedings}{DBLP:conf/models/2011}
\bibitem{DBLP:conf/models/2011}
\bibinfo{editor}{Jon \surnamestart Whittle\surnameend}, \bibinfo{editor}{Tony
  \surnamestart Clark\surnameend} \& \bibinfo{editor}{Thomas \surnamestart
  K{\"u}hne\surnameend}, editors (\bibinfo{year}{2011}):
  \emph{\bibinfo{title}{Model Driven Engineering Languages and Systems, 14th
  International Conference, MODELS 2011, Wellington, New Zealand, October
  16-21, 2011. Proceedings}}. {\sl \bibinfo{series}{Lecture Notes in Computer
  Science}} \bibinfo{volume}{6981}, \bibinfo{publisher}{Springer}.
\newblock \urlprefix\url{http://dx.doi.org/10.1007/978-3-642-24485-8}.

\bibitemdeclare{techreport}{uwe07}
\bibitem{uwe07}
\bibinfo{author}{Uwe \surnamestart Wolter\surnameend} \&
  \bibinfo{author}{Zinovy \surnamestart Diskin\surnameend}:
  \emph{\bibinfo{title}{From Indexed to Fibred Semantics – The Generalized
  Sketch File}}.
\newblock \bibinfo{type}{Technical Report} \bibinfo{number}{361},
  \bibinfo{institution}{Department of Informatics, University of Bergen,
  Norway}.
\newblock
  \bibinfo{note}{{\url{http://www.ii.uib.no/publikasjoner/texrap/pdf/2007-361.pdf}},
  year = 2007,}.

\bibitemdeclare{inproceedings}{Xiong07}
\bibitem{Xiong07}
\bibinfo{author}{Y.~\surnamestart Xiong\surnameend},
  \bibinfo{author}{D.~\surnamestart Liu\surnameend},
  \bibinfo{author}{Z.~\surnamestart Hu\surnameend},
  \bibinfo{author}{H.~\surnamestart Zhao\surnameend},
  \bibinfo{author}{M.~\surnamestart Takeichi\surnameend} \&
  \bibinfo{author}{H.~\surnamestart Mei\surnameend} (\bibinfo{year}{2007}):
  \emph{\bibinfo{title}{Towards automatic model synchronization from model
  transformations}}.
\newblock In: {\sl \bibinfo{booktitle}{ASE}}, pp. \bibinfo{pages}{164--173}.
\newblock \urlprefix\url{http://doi.acm.org/10.1145/1321631.1321657}.

\end{thebibliography}
\appendix
\section{Appendix. Algebra of bidirectional update propagation}
\newcommand\figr{\figref{fig:fbPpg-oper-laws}}

In Section \ref{sect-bx}, we considered operations of update propagation,
but did not specify any laws they must satisfy. Such laws are crucial for
capturing semantics, and the present section aims to specify algebraic laws
for BX. We will do it in an elementary way using tile algebra (rather than
categorically ---  it is non-trivial and left for a future work). We will begin
with the notion of an \emph{alignment \fwk} to formalize delta
composition ($*$ in Section \ref{sect-bx}), and then proceed to algebraic
structures modeling BX --- \emph{symmetric delta lenses}. (Note that the
lenses we will introduce here are different from those defined in
\cite{me-models11}.)

\begin{figure}
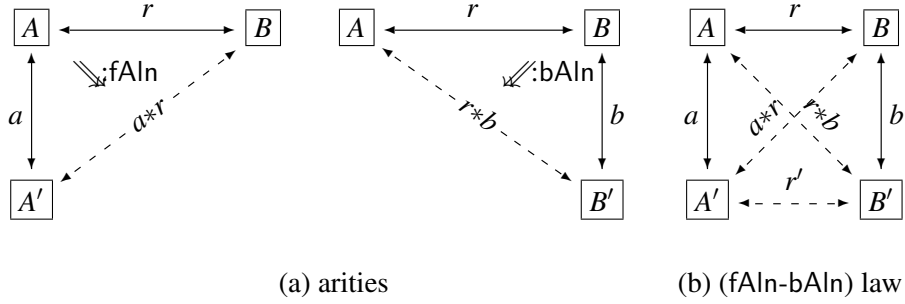

\centering
\begin{tabular}{c@{\qquad}c@{\qquad}c}
\begin{diagram}[w=2em,h=1.5em] 
\dbox{A}&& \rCorrto^r &&\dbox{B}\\%
&\seTilearrow{\falt} &&\ldDercorrto(4,4)~{a{*}r}&\\
\dUpdto<a & && &\\%
&&&&\\
\dbox{A'} & &&&
\end{diagram}%
&
\begin{diagram}[w=2em,h=1.5em] 
\dbox{A} && \rCorrto^r &&\dbox{B}\\%
&\rdDercorrto(4,4)~{r{*}b} &&\swTilearrow{\balt}\quad~~&\\%
& &&&\dUpdto>b \\%
&&&&\\
&&&&\dbox{B'}
\end{diagram}%
&
\begin{diagram} 
\dbox{A} & \rCorrto^r &\dbox{B}\\%
\dUpdto<a %
& %
\ruDercorrto^{a{*}r~~~~~~~}\luDercorrto^{~~~~~~r{*}b}%
& \dUpdto>b \\%
\dbox{A'} & \rDercorrto^{r'} &\dbox{B'}
\end{diagram}%
\\[50pt]
\multicolumn{2}{c}{%
(a)  arities %
} &
(b) (\falt-\balt) law
\end{tabular}
\caption{Realignment operations and their laws \label{fig:align-oper}}%
\end{figure}

\begin{defin}
An \emph{\aln\ \fwk} is given by the following data.

(i) Two categories with pullbacks, \spaA\ and \spaB, called \emph{model
spaces}.  We will consider spans in these categories up to their
equivalence via a head isomorphism commuting with legs. That is, we will
work with equivalence classes of spans, and the term 'span' will refer to an
equivalence class of spans. Then span composition (via pullbacks) is
strictly associative, and we have categories (rather than bicategories) of
spans, $\mathsf{Span_1}(\spaA)$ and $\mathsf{Span_1}(\spaB)$. Their
subcategories consisting of spans with injective legs  will be denoted by
\spaspaA\ and \spaspaB\ resp.

Such spans are to be thought of as \emph{(model) updates}. They will be
depicted by vertical bi-directional arrows, for example, $a$ and $b$ in the
diagrams \figref{fig:align-oper}(a). We will assume that the upper node of
such an arrow is its formal source, and the lower one is the target; the
source is the the original (state of the) model, and the target is the
updated model. Thus, model evolution is directed down.

A span whose upper leg is identity (nothing deleted) is an \emph{insert
update}; it will be denoted by unidirectional arrows going down. Dually, a
span with identity lower leg is a \emph{delete update}; it will be denoted
by a unidirectional arrow going up (but the formal source of such an arrow
is still the upper node).

(ii) For any two objects, $A\in\spaA_0$ and $B\in\spaB_0$, there is a set
$R(A,B)$ of \emph{correspondences} (or \emph{corrs} in short) from
$A$ to $B$. Elements of $R(A, B)$ will be depicted by bi-directional
horizontal arrows, whose formal source is $A$ and the target is $B$.

Updates and corrs will also be called \emph{vertical \emph{and}
horizontal deltas}, resp.

(iii) Two diagram operations over corrs and updates called
\emph{forward} and \emph{backward} \emph{(re)\aln}. Their arities are
shown in \figref{fig:align-oper}(a) (output arrows are dashed). We will
also write $a*r$ for $\falt(a,r)$ and $r*b$ for $\balt(b,r)$. 
We will often skip the prefix 're' and say '\aln' to ease terminology.
\end{defin}

%

\newarrow{Updto}{<}---{>}


There are three laws regulating alignment. Identity updates do not
actually need realignment:
$$
\ide{}{A}*r = r = r*\ide{}{B}%
\leqno (\idaltlaw)
$$
for any corr \bilar{r}{A}{B}.

The result of applying a sequence of interleaving forward and backward
alignments does not depend on the order of application as
shown in \figref{fig:align-oper}(b):%
$$
(a*r)*b = a*(r*b) \leqno (\falt{-}\balt)
$$
for any corr $r$ and any updates $a,b$. 

We will call diagrams like those shown in \figref{fig:align-oper}(a,b)
\emph{commutative} if the arrow at the respective operation output is
indeed equal to that one computed by the operation. For example, diagram
(b) is commutative if $r'=a*r*b$.

Finally, alignment is compositional: for any consecutive updates
\flar{a}{A}{A'}, \flar{a'}{A'}{A''},  \flar{b}{B}{B'}, \flar{b'}{B'}{B''},
the following holds:
$$%
a'*(a*r) = (a;a')*r \mbox{   and   }
(r*b)*b' = r*(b;b') \leqno (\altaltlaw)
$$
where $;$ denotes sequential span composition.

It is easy to see that having an \aln\ \fwk\ amounts to having a functor
\flar{\alpha}{\spaspaA\times\spaspaB}{\setcat}.

\begin{defin}A \emph{symmetric delta lens} (briefly, an sd-lens) is a triple

\noindent $(\alpha,\fppg,\bppg)$ with
\flar{\alpha}{\spaspaA\times\spaspaB}{\setcat} an \aln\ \fwk, and \fppg,
\bppg\ two diagram operations over corrs and updates (called forward and
backward update propagation, resp.). The arities are specified in \figr(a)
with output arrows dashed and output nodes not framed. Sometimes we
will use a linear notation and write $b=a.\fppg(r)$ and $a=b.\bppg(r)$ for
the cases specified in the diagrams.
\end{defin}
{
\cellW=7.25ex%
\cellH=5.0ex
\begin{figure}
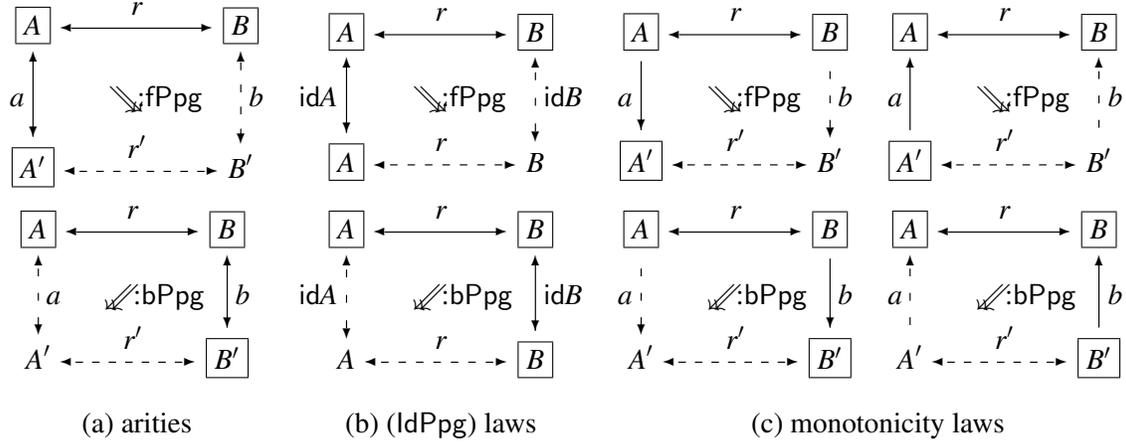

\centering
\begin{tabular}{c@{\quad}c@{\quad}c@{\quad}c}
\begin{diagram}[w=1.1\cellW,h=1.1\cellH] 
\dbox{A} & \rCorrto^r &\dbox{B}\\%
\dCorrto<a & \seTilearrow{\fppg}~~~
            & \dDercorrto>b \\%
\dbox{A'} & \rDercorrto^{r'} & B'
\end{diagram}%
& 
\begin{diagram}[w=\cellW,h=\cellH] 
\dbox{A} & \rCorrto^{r} & \dbox{B}\\%
\dCorrto<{\ide{}{A}} & \seTilearrow{\fppg}~~~
            & \dDercorrto>{\ide{}{B}} \\%
\dbox{A} & \rDercorrto^{r} & B
\end{diagram}%
 &   
 \begin{diagram}[w=\cellW,h=\cellH] 
\dbox{A} & \rCorrto^r &\dbox{B}\\%
\dTo<a & \seTilearrow{\fppg}~~~
            & \dDashto>b \\%
\dbox{A'} & \rDercorrto^{r'} & B'
\end{diagram}%
&     
\begin{diagram}[w=\cellW,h=\cellH] 
\dbox{A} & \rCorrto^r &\dbox{B}\\%
\uTo<a & \seTilearrow{\fppg}~~~
            & \uDashto>b \\%
\dbox{A'} & \rDercorrto^{r'} & B'
\end{diagram}%
\\[30pt]
\begin{diagram}[w=\cellW,h=\cellH] 
\dbox{A} & \rCorrto^r & \dbox{B}\\%
\dDercorrto>a & \swTilearrow{\bppg}~~~
            & \dCorrto>b \\%
A' & \rDercorrto^{r'} & \dbox{B'}
\end{diagram}%
& 
\begin{diagram}[w=\cellW,h=\cellH] 
\dbox{A} & \rCorrto^{r} & \dbox{B}\\%
\dDercorrto<{\ide{}{A}} & \swTilearrow{\bppg}~~~
            & \dCorrto>{\ide{}{B}} \\%
A & \rDercorrto^{r} & \dbox{B}
\end{diagram}%
&   
\begin{diagram}[w=\cellW,h=\cellH] 
\dbox{A} & \rCorrto^r &\dbox{B}\\%
\dDashto<a & \swTilearrow{\bppg}~~~
            & \dTo>b \\%
A' & \rDercorrto^{r'} & \dbox{B'}
\end{diagram}%
&   
   \begin{diagram}[w=\cellW,h=\cellH] 
\dbox{A} & \rCorrto^r &\dbox{B}\\%
\uDashto<a & \swTilearrow{\bppg}~~~
            & \uTo>b \\%
A' & \rDercorrto^{r'} & \dbox{B'}
\end{diagram}%
\\[35pt]
(a) arities &
(b) (\idppglaw)  laws & %
\multicolumn{2}{c}{(c) monotonicity laws}
\end{tabular}
\caption{Operations of update propagation
\label{fig:fbPpg-oper-laws}}%
\end{figure}

}%
Each operation must satisfy the following laws.

\textbf{\emph{Stability}} or \idppglaw\ law: if nothing changes on one
side, nothing happens on the other side as well, that is, identity mappings
are propagated into identity mappings as shown by diagrams \figr(b).

\textbf{\emph{Monotonicity}}: Insert updates are propagated into
inserts, and delete updates are propagated into deletes, as specified in
\figr(c).

\textbf{\emph{Monotonic Compositionality}} or \ppgppglaw\ law:
composition of two consecutive inserts is propagated into composition of
propagations as shown by the left diagram in \figref{fig:monot-PpgPpg}
(to be read as follows: if the two squares are \fppg, then the outer
rectangle is \fppg \ as well). The right diagram  specifies compositionality
for deletes. The same laws are formulated for \bppg.
\newarrow{Updto}----{>}
\newarrow{Derupdto}{}{dash}{}{dash}{>}

\begin{wrapfigure}{R}{0.45\textwidth}
\vspace{-0.1cm}%
\begin{tabular}{c@{\qquad}c} 
    \begin{diagram}[w=\cellW,h=\cellH] 
\dbox{A} & \rCorrto^r &\dbox{B}\\%
\dUpdto<a & \seTilearrow{\fppg}~~~
            & \dDerupdto>b \\%
\dbox{A'} & \rDercorrto^{r'} & B'  \\%
\dUpdto<{a'} & \seTilearrow{\fppg}~~~
            & \dDerupdto>b'\\%
            \dbox{A''} & \rDercorrto^{r''} & B''%
             &&\\
\end{diagram}%
  & %
    \begin{diagram}[w=\cellW,h=\cellH] 
\dbox{A} & \rCorrto^r &\dbox{B}\\%
\uUpdto<a & \seTilearrow{\fppg}~~~
            & \uDerupdto>b \\%
\dbox{A'} & \rDercorrto^{r'} & B'  \\%
\uUpdto<{a'} & \seTilearrow{\fppg}~~~
            & \uDerupdto>b'\\%
            \dbox{A''} & \rDercorrto^{r''} & B''%
            &&\\
\end{diagram}%
\end{tabular}
\caption{Monotonic (\ppgppglaw) laws \label{fig:monot-PpgPpg}}%
\vspace{-0.1cm}
\end{wrapfigure}

Note that we do  not require compositionality for propagation of general
span updates. The point is that interleaving inserts and deletes can
annihilate, and lost information cannot be restored: see \cite{Foster07,
me-jot11,me-models11} for examples.

\textbf{\emph{Commutativity}}: Diagrams \figr(a) must be commutative
in the sense that $a*r*b = r'$.

\begin{figure}
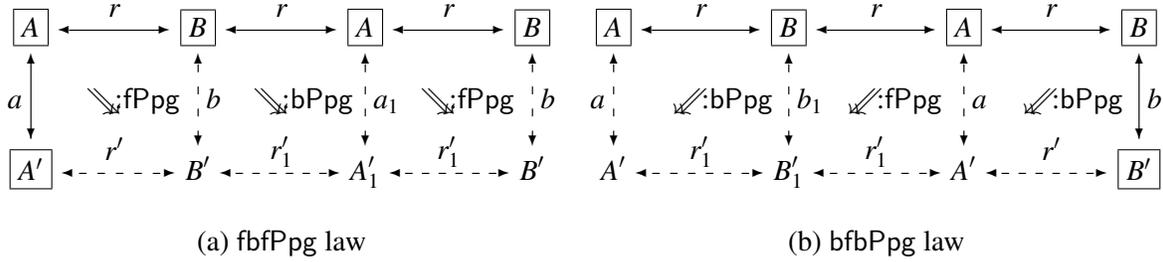

\centering
\begin{tabular}{c@\quad c}
\begin{diagram}[w=1.05\cellW,h=1.1\cellH] 
\dbox{A} & \rCorrto^r     & \dbox{B} & \rCorrto^{r} & \dbox{A}%
               &  \rCorrto^{r} & \dbox{B}
\\%
\dCorrto<a & \seTilearrow{\fppg}~~~ & \dDercorrto>b %
                 & \seTilearrow{\bppg}~~~ & \dDercorrto>{a_1}
                  & \seTilearrow{\fppg}~~~ & \dDercorrto>b
\\%
\dbox{A'} & \rDercorrto^{r'}  & %
B'            & \rDercorrto^{r_1'} & A_1' %
               & \rDercorrto^{r_1'} & B'
\end{diagram}%
&    %
\begin{diagram}[w=1.1\cellW,h=1.1\cellH] 
\dbox{A} & \rCorrto^r     & \dbox{B} & \rCorrto^{r} & \dbox{A}%
               &  \rCorrto^{r} & \dbox{B}
\\%
\dDercorrto<a & \swTilearrow{\bppg}~~~ & \dDercorrto>{b_1} %
                 & \swTilearrow{\fppg}~~~ & \dDercorrto>{a}
                  & \swTilearrow{\bppg}~~~ & \dCorrto>b
\\%
{A'} & \rDercorrto^{r_1'}  & %
 B_1'            & \rDercorrto^{r_1'} & A' %
               & \rDercorrto^{r'} & \dbox{B'}
\end{diagram}%
\\[40pt]
(a) \fbfppglaw\ law  & (b) \bfbppglaw\ law
\end{tabular}
\caption{Round-tripping laws. (Scenario in diagram (b) ``runs'' from the right to the left.)
\label{fig:invert-laws}}%
\end{figure}

Finally, forward and backward propagation must be coordinated with each
other by some \textbf{\emph{invertibility}} law. Given a corr
\flar{r}{A}{B}, an update \flar{a}{A}{A'} is propagated into update
$b=a.\fppg(r)$, which can be propagated back to update $a'=b.\bppg(r)$.
For an ideal situation of \emph{strong invertibility}, we should require
$a'=a$. Unfortunately, this does not hold in general because the
\spaspaA-specific part of the information is lost in passing from $a$ to
$b$, and cannot be restored \cite{me-models11}. However, it makes
sense to require the following  {\emph{weak invertibility}} specified in
\figref{fig:invert-laws}, which does hold in a majority of practically
interesting situations, \eg, for BX determined by TGG-rules
\cite{frank-models11}. The law \fbfppglaw\  says that although
$a_1=a.\fppg(r).\bppg(r)\neq a$, $a_1$ is equivalent to $a$ in the sense
that $a_1.\fppg(r) = a.\fppg(r)$. Similarly for the \bfbppglaw\ law.

\bigskip
The notion of sd-lens is specified above in elementary terms using tile
algebra. Its categorical underpinning is not evident, and we only present
several brief remarks.

1) An \aln\ \fwk\  \flar{\alpha}{\spaspaA\times\spaspaB}{\setcat} can be
seen as a profunctor, if \spaA-arrows will be considered directed up (i.e.,
the formal source of update $a$ in diagram \figref{fig:align-oper}(a) is
$A'$, and the target is $A$). Then \aln\ amounts to a functor
\flar{\alpha}{{\spaspaA}^{\sf{op}}\times\spaspaB}{\setcat}, that is, a
profunctor \proflar{\alpha}{\spaspaB}{\spaspaA}. Note that reversing
arrows in \spaspaA\ actually changes the arity of operation \falt: now its
input is a pair $(a,r)$ with $a$ an update and $r$ a corr from the target of
$a$, and the output is a corr $r'$ from the source of $a$, that is, re\aln\
goes back in time.

2) Recall that operations \fppg\ and \bppg\ are functorial wrt. injective
arrows in \spaA, \spaB, not wrt. arrows in \spaspaA, \spaspaB. However, if
we try to resort to \spaA, \spaB\ entirely and define \aln\ wrt. arrows in
\spaA, \spaB, then we will need two \falt\ operations with different arities
for inserts and deletes, and two \balt\ operations with different arities for
inserts and deletes. We will then have four functors
\flar{\alpha_i}{\spaA\times\spaB}{\setcat} with $i$ ranging over
four-element set $\{insert, delete\}\times \{\spaA, \spaB\}$.

3)  The weak invertibility laws suggest that a Galois connection/adjunction
is somehow hidden in sd-lenses.

4) Working with chosen spans and pullbacks rather than with their
equivalence classes provides a more constructive setting (given we assume
the axiom of choice), but then associativity of span composition only holds
up to chosen natural isomorphisms, and  \spaspaA\ and \spaspaB\ have to
be considered bicategories rather than categories.

All in all, we hope that the categorical analysis of asymmetric delta lenses
developed by Johnson \etal\ \cite{joro10,joro12} could be extended to
capture the symmetric case too.

\end{document}